\documentstyle[aps,prd,preprint,amssymb,eqsecnum,tighten,epsf]{revtex}
\begin{document}
\draft

\title{
\rightline{\small{\em Phys. Rev. D {\bf 63}, 124013 (2001)}}
The close limit from a null point of view: the advanced solution}
\author{
Manuela Campanelli$^{1}$,
Roberto G\'omez$^{2}$,
Sascha Husa$^{1,2}$,
Jeffrey Winicour$^{1,2}$, and \\
Yosef Zlochower$^{2}$
}

\address{1. Max-Planck-Institut f\" ur Gravitationsphysik,
Albert-Einstein-Institut, 14476 Golm, Germany.}
\address{2. Department of Physics and Astronomy, University of Pittsburgh,
Pittsburgh, Pennsylvania 15260.}

\maketitle
\begin{abstract}

We present a characteristic algorithm for computing  the perturbation of a
Schwarzschild spacetime by means of solving the Teukolsky equation. We
implement the algorithm as a characteristic evolution code and apply it to
compute the advanced solution to a black hole collision in the close
approximation. The code successfully tracks the initial burst and  quasinormal
decay of a black hole perturbation through 10 orders of magnitude and tracks
the final power law decay through an additional 6 orders of magnitude.
Determination of the advanced solution, in which ingoing radiation is absorbed
by the black hole but no outgoing radiation is emitted, is the first stage of a
two stage approach to determining the retarded solution, which provides the
close approximation waveform with the physically appropriate boundary condition
of no ingoing radiation.

\end{abstract}


\section{Introduction}

In this work, we present the advanced solution for a perturbation of a
Schwarzschild background describing the head-on collision of black holes in the
close approximation where the merger takes place in the far past. We compute
the solution by means of a characteristic evolution of the Weyl tensor, as
governed by the Teukolsky equation~\cite{Teukolsky72,Teukolsky73}. The advanced
solution corresponds to Stage I of a new two stage approach to the vacuum
binary black hole problem~\cite{kyoto}. In subsequent work (Stage II), we will
use the results of this first stage to compute the physically more relevant
retarded solution. This perturbative solution will in turn provide a valuable
reference point for the physical understanding of a fully nonlinear treatment
of binary black holes being pursued by a similar two stage
strategy~\cite{ndata}, a computationally feasible problem using an existing
characteristic code~\cite{high}. The perturbative results also provide a new
perspective on the physical picture previously obtained by applying the Cauchy
problem to the close approximation~\cite{pp}, especially with regard to global
issues which have not yet been explored in the Cauchy formulation.

In the characteristic formulation of this problem, the advanced solution is
simpler to compute than the retarded solution because of the global
relationship between the null hypersurfaces on which boundary information is
known. One of these hypersurfaces is the black hole event horizon ${\cal H}^+$
whose perturbation must correspond to the close approximation of a binary black
hole. In the retarded problem, the other null hypersurface (in a conformally
compactified description) is past null infinity ${\cal I}^-$ where the incoming
radiation must vanish. Because ${\cal H}^+$ and ${\cal I}^-$ are disjoint,
there is no direct way to use data on those two hypersurfaces to pose a
characteristic initial value problem.

However, in the advanced problem, it is at future null infinity  that the
radiation is required to vanish. Since ${\cal H}^+$ and ${\cal I}^+$ formally
intersect at future time infinity $I^+$, they can be used to pose a
characteristic initial value problem to evolve backward in time and compute the
exterior region of spacetime. Potential difficulties in dealing with $I^+$ are
avoided by posing the no outgoing radiation condition on an ingoing null
hypersurface $J^+$ which approximates ${\cal I}^+$ and intersects ${\cal H}^+$
at a late time  when the perturbation of the black hole has effectively died
out. These data on ${\cal H}^+$ and $J^+$ then constitute a standard
double-null initial value problem~\cite{sachsdn,haywdn,fried81,rend} for the
exterior spacetime, in which ingoing radiation is absorbed by the black hole
but there is no outgoing radiation.

This advanced solution provides the radiation incident from ${\cal I}^-$. In
Stage II of the approach, this ingoing radiation will be used to generate a
``source free'' {\em advanced minus retarded} solution. A purely retarded
solution can then be produced by superposition with the Stage I solution.
Although we do not address Stage II in this paper, we will discuss the role of
time reflection symmetry in the perturbation equations, which simplifies the
technical details in carrying out the superposition. From a time reversed point
of view, the Stage I solution is equivalent to the retarded solution for a
``head-on'' white hole fission, with the physically relevant boundary
conditions that radiation is emitted but that there is no incoming radiation
from ${\cal I}^-$. It is convenient here to formulate the Stage I results in
terms of such a white hole fission since the characteristic evolution then
takes the standard form of being carried out forward in retarded time.

The  close approximation has been extremely useful for testing fully
nonlinear Cauchy evolution codes. The results of numerical Cauchy evolution and
close-limit perturbative theory are in excellent agreement in the appropriate
regime, giving great confidence in both approaches~\cite{pp,Baker99a}.
Furthermore, the perturbative approach provides an important tool for the
interpretation and physical understanding of those
results~\cite{Baker2000b}.

Clearly, this vital synergism between numerical and perturbative approaches
should also extend to characteristic evolution. However, in all perturbative
studies performed to date, the  background geometry has either been the
Schwarzschild spacetime in standard Schwarzschild coordinates or the Kerr
spacetime in Boyer-Lindquist coordinates. These coordinate systems are
appropriate for comparison with results from nonlinear Cauchy evolution but, to
our knowledge, there does not yet exist a treatment of the close approximation
in terms of null coordinates appropriate for the comparison with nonlinear
characteristic evolution.

This work provides such a framework. The methods and results presented here are
expected to have direct bearing on the study of binary black holes presently
underway using a fully nonlinear characteristic code. Characteristic evolution
has been totally successful in evolving 3-dimensional single black
hole spacetimes for effectively infinite times ($t\approx 60,000 M$ in terms of
black hole mass) \cite{Gomez98a}. Although it is not yet known to what extent
the characteristic approach can handle the inspiral and merger of binary black
holes, it is clear that the limitations are due to difficulties in treating
caustics and not due to high nonlinearity.

The characteristic data which has been obtained for the nonlinear description
of a binary black hole spacetime generate close approximation data for a
perturbative solution. We present here a numerical code to evolve such data as
a perturbation of a Schwarzschild spacetime in null coordinates.  Fortunately
for our purposes, the perturbative formalism due to
Teukolsky~\cite{Teukolsky73} is amenable to a reasonably straightforward change
of background coordinates, as observed in Ref.~\cite{Campanelli2000a}.

The Teukolsky equation is based upon the decomposition of the Einstein
equations and Bianchi identities  in terms of a conveniently chosen complex
null tetrad, as carried out by Newman and Penrose~\cite{np} in the early
1960's. The Newman-Penrose formalism allowed Teukolsky to construct
a single master wave equation for the perturbations of the Kerr metric in terms
of the Weyl curvature components $\psi_4$ (describing outgoing radiation)  or
$\psi_0$ (describing ingoing radiation). The Teukolsky formalism provides a
completely gauge invariant spherical harmonic multipole decomposition for both
even and odd parity perturbations in terms of radial wave equations. For a Kerr
black hole with angular momentum, there is no similar multipole decomposition
of metric  perturbations in the time domain (as opposed to the frequency
domain of Fourier modes). In the non-rotating case, the Teukolsky
equation reduces to the so-called Bardeen-Press equation~\cite{Bardeen73a}

Since the 1970's the Teukolsky equation for the first order perturbations
around a rotating black hole has been Fourier transformed and integrated in the
frequency domain for a variety of situations where initial data played no
role~\cite{Chandrasekhar83,FN89}.  In order to avoid the important but
difficult problem of prescribing physically appropriate initial data for that
equation, the computation of gravitational radiation has been restricted to the
cases of unbounded particle trajectories or circular motion. The first
evolution code to integrate the Teukolsky  equation in the time domain, in
Boyer-Lindquist coordinates,  was recently developed \cite{Krivan97a} and
successfully tested~\cite{Campanelli98b}. In order to incorporate initial data
representing realistic  astrophysical initial data for the late stage of binary
black  hole coalescence, $3+1$ expressions connecting  $\psi_4$ and its time
derivative to Cauchy data (3-metric $h_{ij}$ and extrinsic curvature $K_{ij}$)
satisfying the Hamiltonian and momentum constraints
$$\psi_4=\psi_4(h_{ij},K_{ij}),~~ \partial_t\psi_4=\dot\psi_4(h_{ij},K_{ij}),$$
have been worked out
explicitly~\cite{Campanelli98a,Campanelli98b,Campanelli98c,Campanelli99}.

In Sec.~\ref{sec:pert}, we  specify a null background tetrad suitable to
re-express the Teukolsky equations for $\psi_0$ and $\psi_4$ in null
coordinates appropriate for characteristic evolution. We also discuss various
global aspects of these equations which are important for numerical evolution.
In  Sec.~\ref{sec:data}, we discuss data for the Teukolsky equation. We present
null data for linearized Robinson-Trautman solutions, which provide an analytic
check on numerical accuracy, and null data for the close approximation to a
white hole fission. In Sec.~\ref{sec:algorithm}, we discuss the numerical
algorithm used to carry out the characteristic evolution. The properties of
the close approximation waveforms, are presented in
Sec.~\ref{sec:waveforms}.

Notation and Conventions: We use a metric of signature $(-+++)$ and
a null tetrad with  normalization $l^a n_a=-m^a\bar{m}_a=-1$, so that
$g_{ab}=2(m_{(a}\bar{m}_{b)}-l_{(a} n_{b)})$.
We use $q_{AB}$ to represent the standard unit sphere metric in angular
coordinates $x^A=(\vartheta,\varphi)$ and set $q^{AB} = q^{(A}\bar q^{B)}$,
where $q^{AB}q_{BC}=\delta^A_C$, with $q^A=(1,i/\sin\vartheta)$.
We use $q^A$ to define the $\eth$ operator with the
convention $\eth f =q^A \partial_A f$, for a spin-weight 0 function $f$.
Complex conjugation is denoted with a ``bar'', e.g. $\Re(f) =(f+\bar f)/2$.
The conventions of the present paper result in a different form of
the perturbation equations from that originally given by Teukolsky.

\section{The perturbation equations}
\label{sec:pert}

\subsection{The background Schwarzschild metric in outgoing horizon
coordinates}

The Schwarzschild metric in standard coordinates is
\begin{equation}
     ds^2= -(1-\frac{2 M}{r})dt^2+dr^2(1-\frac{2 M}{r})^{-1}
                +r^2 q_{AB}dx^Adx^B.
\end{equation}
In outgoing Eddington-Finkelstein coordinates, where $\tilde{u}=t-r^*$  is a
null coordinate and $r^*=r+2M\log(\frac{r}{2M}-1)$, the Schwarzschild  metric
takes the
Bondi form
\begin{equation}
   ds^2  = -(1-\frac{2 M}{r})d\tilde{u}^2-2d\tilde{u}dr+r^2 q_{AB}dx^Adx^B.
   \label{eq:outEF}
\end{equation}
These coordinates specialize to spherical symmetry the general procedure for
constructing a Bondi null coordinate system~\cite{bondi}. They patch two
quadrants of the Kruskal manifold: the exterior spacetime quadrant and the
quadrant following the initial singularity.

For the anticipated comparison with a fully nonlinear description of a white
hole, it is useful to introduce another null coordinate system based upon the
affine parameter $u=-M e^{-\tilde{u}/4M}$ along the ingoing null hypersurface
$r=2M$ that forms the white hole horizon. We set $u=0$ at the intersection of
the black hole and white hole horizons, i.e. at the $r=2M$ bifurcation sphere
(corresponding to $\tilde{u}=+\infty$ in Eddington-Finkelstein coordinates).
The metric then becomes
\begin{equation}
      ds^2  = -(1-\frac{2 M}{r})\frac{16 M^2}{u^2}du^2+ \frac{8 M}{u}du dr
           +r^2 q_{AB}dx^Adx^B .
\end{equation}
In addition, we introduce an affine parameter $\lambda$ along the outgoing null
geodesics in the $r$ direction, with the affine freedom fixed by requiring that
$\lambda=0$ when $r=2M$ and that $g^{ab}(\partial_a u) \partial_b \lambda=-1$.
This implies  $$\lambda = -\frac{4 M(r-2M)}{u}.$$

In these $(u,\lambda)$ coordinates, the metric takes the form
\begin{equation}
    ds^2  = -W du^2-2dud\lambda  +r^2 q_{AB}dx^Adx^B ,
\label{eq:amet}
\end{equation}
where
\begin{equation}
         r=2M -\frac {\lambda u}{4M}
\label{eq:r}
\end{equation}
and
\begin{equation}
     W=\frac{2\lambda^2}{\lambda u -8M^2}.
\end{equation}
These coordinates specialize to spherical symmetry the general procedure for
constructing a Sachs null coordinate system designed for the double-null
initial value problem~\cite{sachsdn}. For this reason, they are especially
useful for the study of horizons in the nonlinear regime. They are also useful
in the perturbative regime because they cover the entire Kruskal manifold $r>0$
with remarkably simple analytic behavior, as first discovered by
Israel~\cite{israel}. Since $g^{\lambda\lambda}=W=-\lambda^2/(2Mr)$ the
hypersurfaces $\lambda=const$ are everywhere spacelike (so that the
$u$-direction is spacelike) except on the white hole horizon where $\lambda =0$
and the $u$-direction is null. The surfaces $u=const$ are everywhere null.  The
spacelike surfaces $\lambda=const >0$ ($\lambda=const <0$) can be used as
partial Cauchy hypersurfaces to cover the two quadrants above  (below)
$\lambda=0$ in the Kruskal manifold.

\subsection{The Teukolsky equations}

By aligning  a complex null tetrad, $(l^{\mu},n^{\mu},m^{\mu},\bar{m}^{\mu})$
with the degenerate principal null directions  of a Petrov type D
background spacetime,  Teukolsky \cite{Teukolsky73} was able to express
the vacuum perturbation equations for the Weyl curvature scalars
$\psi_0=C_{abcd}l^a m^b l^c m^d$ (of spin-weight $s=2$) and
$\psi_4=C_{abcd}n^{a}\bar{m}^{b}n^{c}\bar{m}^{d}$ (of spin-weight $s=-2$)
of the Newman-Penrose (NP) formalism \cite{Newman62a} as the
simple wave equations

\begin{eqnarray}
\left[\left(D+3\epsilon-\bar{\epsilon}+4\rho+\bar{\rho}\right)
\left(\Delta+4\gamma-\mu\right)-\left(\delta-\bar{\pi}+\bar{\alpha}
+3\beta+4\tau\right)\left(\bar{\delta}-\pi+4\alpha\right)
-3\psi_2\right]\psi_0=0, \label{teukpsi0}
\end{eqnarray}
\begin{eqnarray}
\left[\left(\Delta-3\gamma+\bar{\gamma}-4 \mu-\bar{\mu}\right)
\left(D-4\epsilon+\rho\right)-\left(\bar{\delta}+\bar{\tau}-
\bar{\beta}-3\alpha-4\pi\right)\left(\delta+\tau-4\beta\right)
-3\psi_2\right]\psi_4=0. \label{teukpsi4}
\end{eqnarray}
Here the spin coefficients
$\alpha=(l_{a;b}n^{a}\bar{m}^{b}-m_{a;b}\bar{m}^{a}
\bar{m}^{b})/2$, $\beta=(l_{a;b}n^{a}m^{b}-m_{a;b}
\bar{m}^{a}m^{b})/2$, $\gamma=(l_{a;b}n^{a}n^{b}-
m_{a;b}\bar{m}^{a}n^{b})/2$, $\epsilon=(l_{a;b}n^{a}
l^{b}-m_{a;b}\bar{m}^{a}l^{b})/2$, $\tau=l_{a;b}m^{a}
n^{b}$, $\pi=-n_{a;b}\bar{m}^{a}l^{b}$, $\rho=l_{a;b}m^{a}
\bar{m}^{b}$ and $\mu=-n_{a;b}\bar{m}^{a}m^{b}$
are computed using the background geometry and the directional
derivatives are $D=l^{a}\partial_{a}$, $\Delta=n^{a}\partial_{a}$,
$\delta=m^{a}\partial_{a}$ and $\bar{\delta}=\bar{m}^{a}\partial_{a}$.
The Weyl scalars $\psi_0$ or $\psi_4$ are first order quantities
in perturbation theory while
$\psi_2=C_{abcd}l^{a}m^{b}\bar{m}^{c}n^{d}=-M/r^3$ is a
zeroth order curvature quantity.

This formulation has several advantageous features: (i) It is a completely
first order gauge invariant description (i.e.the perturbative
Weyl scalars $\psi_0$ or $\psi_4$ are invariant not only under
infinitesimal coordinate transformations but also under null
rotations of the tetrad); (ii) It does not rely on any frequency or
multipole decomposition (i.e. the above equations can be directly
integrated in the time domain); (iii) The Weyl scalars are objects
defined in the full nonlinear theory, where $\psi_0$ can be prescribed
as constraint-free data on an outgoing null hypersurface and $\psi_4$
as constraint-free data on an ingoing null hypersurface.

Since Eq's. (\ref{teukpsi0})-(\ref{teukpsi4}) are expressed
in a covariant form, it is straightforward to write them
explicitly in any background coordinate system.
Specializing thus to the Schwarzschild background metric
in the Israel coordinates introduced in
Eq.~(\ref{eq:amet}), we choose a null tetrad
\begin{eqnarray}
  l^a&=&-\nabla^a u = (\frac{\partial}{\partial\lambda})^a =
  \left[0,1,0,0\right],\nonumber\\
    n^a&=&(\frac{\partial}{\partial u})^a -
       \frac{W}{2} (\frac{\partial}{\partial\lambda})^a =
   \left[1,\frac{\lambda^2}{4Mr},0,0\right],\nonumber\\
   m^a&=&\frac{1}{{\sqrt 2} r} q^a,\nonumber\\
     \bar{m}^a&=&\frac{1}{{\sqrt 2} r} \bar q^a,
\label{NPtetrad}
\end{eqnarray}
where
\begin{equation}
         q^a = (\frac{\partial}{\partial\vartheta})^a +
        \, \frac{i}{\sin\vartheta}(\frac{\partial}{\partial\varphi})^a=
          \left[0,0,q^A \right].
\end{equation}

Correspondingly, the only non-vanishing background NP quantities are,
\begin{eqnarray}
\alpha&=&-\beta=\frac{cot(\vartheta)}{2\sqrt{2}r},~~~~
\gamma=\frac{(r+2M)\lambda}{8Mr^2},~~~~
\mu=\frac{\lambda}{2r^2},~~~~
\rho=-\frac{u}{4Mr},\nonumber\\
D&=&\frac{\partial}{\partial{\lambda}},~~~~
\Delta=\frac{\partial}{\partial{u}}
+\frac{\lambda^2}{4Mr}\frac{\partial}{\partial\lambda},~~~~
\delta=\frac{1}{\sqrt{2}r}\left(\frac{\partial}{\partial{\vartheta}}
+\frac{i}{\sin\vartheta}\frac{\partial}{\partial{\varphi}}\right),~~~~
\bar{\delta}=\frac{1}{\sqrt{2}r}\left(\frac{\partial}{\partial{\vartheta}}
-\frac{i}{\sin\vartheta}\frac{\partial}
{\partial{\varphi}}\right).\label{NPquantities}
\end{eqnarray}
Substitution of the above NP quantities into Eq's. (\ref{teukpsi0})
-(\ref{teukpsi4}) reduces the Teukolsky equations to
\begin{equation}
(L_0 + \frac {L^2}{2r^2}) \psi_0 = 0, ~~~{\rm and}~~~
(L_4 +\frac {L^2}{2r^2} ) \psi_4 = 0, \label{teukopsi}
\end{equation}
where
\begin{eqnarray}
  L_0&=&\frac{1}{4Mr}\lambda^2\partial^2_\lambda+
    \partial_u\partial_\lambda-\frac{5}{4Mr}u\partial_u
   -\frac{1}{2Mr^2}\lambda(3M-4r)\partial_\lambda
  +\frac{5}{2 r M}, \nonumber \\
  L_4 &=&\frac{1}{4Mr}\lambda^2\partial^2_\lambda+
   \partial_u\partial_\lambda-\frac{1}{4Mr}u\partial_u
   -\frac{7}{2r^2}\lambda\partial_\lambda
    -\frac{r^2 - 16 M^2 + 4 M r}{2 M r^3}
\label{teukops}
\end{eqnarray}
and $L^2=-\bar \eth \eth$ is the angular momentum squared operator. In order to
treat the radiation near ${\cal I}^+$ it is also advantageous to consider a
boosted tetrad  $({\tilde l}^a, {\tilde n}^a, m^a, \bar m^a )$, with ${\tilde
l}^a=-\nabla^{a} \tilde{u}$ and ${\tilde n}^a$ satisfying ${\tilde l}^a {\tilde
n}_a = -1$. We accordingly define boosted Weyl scalars ${\tilde \psi}_0 =
C_{abcd} {\tilde l}^a m^b {\tilde l}^c m^d$ and ${\tilde \psi}_4 = C_{abcd}
{\tilde n}^a \bar m^b {\tilde n}^c \bar m^d$. This boosted tetrad is adapted to
the affine time $\tilde u$ at ${\cal I}^+$ rather than the affine time $u$
at the horizon.

\subsection{Spin-weight-zero versions of the Teukolsky equations}

It is useful to convert Eq's. (\ref{teukopsi}) - (\ref{teukops}) into
spin-weight-zero equations. Considering  the commutation relation $(\bar \eth
\eth -\eth \bar \eth)\eta =2s\eta$ for a spin-weight $s$ function $\eta$, and
setting $\psi_0 =\eth^2 \Phi_0$ so that $\bar \eth\eth\psi_0 = \eth^2
(6+\bar\eth\eth)\Phi_0$, the Teukolsky equation for $\psi_0$ becomes, after
factoring out an overall $\eth^2$,
\begin{equation}
\left[\frac{1}{4Mr}\lambda^2\partial^2_\lambda+
\partial_u\partial_\lambda-\frac{5}{4Mr}u\partial_u
-\frac{1}{2Mr^2}\lambda(3M-4r)\partial_\lambda
+\frac{5}{2 r M} - \frac{(6+\bar \eth \eth)}
{\,2 r^2}\right]\Phi_0=0.
\label{teukopsi0s0}
\end{equation}
Similarly, setting $\psi_4 =\bar \eth^2 \Phi_4$  so that
$\bar \eth\eth\psi_4 = \bar\eth^2 (2+\bar\eth\eth)\Phi_4$ and
after factoring out an overall $\bar\eth^2$, the Teukolsky equation
for $\Phi_4$  takes the spin-weight-zero form
\begin{equation}
\left[\frac{1}{4Mr}\lambda^2\partial^2_\lambda+
\partial_u\partial_\lambda-\frac{1}{4Mr}u\partial_u
-\frac{7}{2r^2}\lambda\partial_\lambda -
\frac{r^2 - 16 M^2 + 4 M r}{2 M r^3}-
\frac{(2+\bar\eth \eth)}{2\,r^2}\right]\Phi_4=0.
\label{teukopsi4s4}
\end{equation}

These equations can be re-expressed in terms of the Laplacian
\begin{equation}
       D^2  = -\frac{2 \lambda ( 16 M^2 - u \lambda)}{(8 M^2 - u \lambda)^2}
          \partial_\lambda -2 \partial_u \partial_\lambda
         -\frac{2 \lambda^2}{8 M^2 - u \lambda} \partial_\lambda^2,
\label{D}
\end{equation}
defined by the metric $d\bar s^2 = -W du^2 - 2 du d\lambda$ induced by the
Schwarzschild metric on the
2-dimensional $(u,\lambda)$ subspace. Discussion of the
asymptotic behavior of the Weyl tensor is most convenient in terms of the
variables $F_0 = r^5\Phi_0$ and $F_4 = r\Phi_4$. The spin-weight-zero Eq's
(\ref{teukopsi0s0}) and (\ref{teukopsi4s4}) then reduce to
\begin{eqnarray}
\label{eq:F_0}
\left (D^2 + T_0 \right) F_0 &=& 0 \\
\label{eq:F_4}
\left (D^2 + T_4 \right) F_4 &=& 0,
\end{eqnarray}
where
$$
T_0=  -\frac{(6 M+r) \lambda}{M r^2}\partial_\lambda - \frac{30 M}{r^3}
+ \frac{(6+\bar \eth \eth)} {\,r^2}
$$
and
$$
T_4 =  \frac{(6 M+r) \lambda}{M r^2} \partial_\lambda - \frac {6 M}{r^3}
 + \frac{(2+\bar\eth \eth)}{\,r^2}
$$
do not contain $\partial_u$ terms.

Asymptotic flatness requires that $F_0$ and $F_4$ have finite limits at ${\cal
I}^+$. This is consistent with the asymptotic forms of Eq's (\ref{eq:F_0}) and
(\ref{eq:F_4}), which asymptote to one-dimensional wave equations for solutions
whose radial derivative falls off uniformly as $O(1/r^2)$. The limit of $F_4$
determines the outgoing gravitational radiation waveform and the limit of $F_0$
is related to the retarded quadrupole moment of the system. More precisely,
$\lim r\tilde \psi^0_4 =\partial_{\tilde u} \bar N$, where $N$ is the standard
definition of the Bondi news function.

In the limit $\lambda \to \infty$,
     $$\frac{(6 M+r) \lambda}{M r^2} \to -\frac{4}{u}, $$
so that the appearance of this term in $T_0$ and $T_4$ can cause
inaccuracy in the numerical solution of Eq's (\ref{eq:F_0}) and
(\ref{eq:F_4}) at late Bondi times on ${\cal I}^+$ (as $-u\rightarrow
0$). In particular, the late time behavior of the radiation waveform,
can be more accurately computed by an evolution algorithm for the Weyl
component $\tilde \psi_4$. Setting $\tilde\psi_0 =\eth^2 \tilde\Phi_0$
and $\tilde \psi_4 =\bar \eth^2 \tilde \Phi_4$, the fields $\tilde F_0
=r^5\tilde \Phi_0$ and $\tilde F_4 =r\tilde \Phi_4$ satisfy

\begin{eqnarray}
\label{eq:tildF_0}
           \left (D^2 + S_0 \right) \tilde F_0 &=& 0, \\
\label{eq:tildF_4}
               \left (D^2 + S_4 \right) \tilde F_4 &=& 0,
\end{eqnarray}
where
$$
    S_0=  \frac{16M(r-3M)}{ur^2}\partial_\lambda - \frac{30 M}{r^3}
    + \frac{(6+\bar \eth \eth)} {\,r^2}
    = -\frac{4(r-3M)}{ r^2} \partial_r - \frac{30 M}{r^3}
    + \frac{(6+\bar \eth \eth)} {\,r^2}
    \label{eq:S0}
$$
and
$$
  S_4 =  -\frac{16M(r-3M)}{u r^2} \partial_\lambda - \frac {6 M}{r^3}
      + \frac{(2+\bar\eth \eth)}{\,r^2}
   = \frac{4(r-3M)}{ r^2} \partial_r - \frac {6 M}{r^3}
      + \frac{(2+\bar\eth \eth)}{\,r^2}.
      \label{eq:S4}
$$
In these variables, the deviation of the Teukolsky equations
(\ref{eq:tildF_0}) and (\ref{eq:tildF_4}) from a 1-dimensional wave equation is
independent of time at a fixed $r$. These equations are well behaved at ${\cal
I}^+$, i.e. after a compactified coordinate such as $x=1/r$ is introduced.
Because $\tilde F_4 = (u/4M)^2 F_4$, where $F_4$ must be regular throughout the
Kruskal manifold (since it is constructed with a regular basis), it follows
that $\tilde F_4 \rightarrow 0$ as the black hole horizon is approached. This
facilitates an accurate long term evolution of the waveform using Eq.
(\ref{eq:tildF_4}).

Note also, however, that $\tilde F_0 = (4M/u)^2 F_0$ so that $\tilde F_0$ is
singular on the black hole horizon and thus a poor choice of variable for long
term evolution. The opposite signs of the coefficients of $\partial_r$ in
$S_0$ and $S_4$ are responsible for this behavior, as can be seen by ignoring
the remaining potential terms and freezing the coefficient of $\partial_r$ at
$r=2M$, so that Eq's (\ref{eq:tildF_0}) and (\ref{eq:tildF_4}) reduce to
\begin{eqnarray}
\label{eq:wave0}
  (2\partial_{\tilde u} -\partial_r -\frac{1}{M})\partial_r\tilde F_0 &=& 0, \\
\label{eq:wave4}
    (2\partial_{\tilde u} -\partial_r +\frac{1}{M})\partial_r\tilde F_4 &=& 0,
\end{eqnarray}
in terms of retarded Bondi coordinates. In this approximation, both of these
equations admit purely outgoing waves $\tilde F(\tilde u)$. However, an ingoing
$\tilde F_0$ wave has the exponentially singular behavior
   $$ \tilde F_0 = f(\tilde u +2r)e^{\frac {\tilde u - \tilde u_0}{2M}} $$
as an initial pulse $f(\tilde u_0 +2r)$ approaches the black hole horizon as
$\tilde u \rightarrow \infty$. In contrast, an ingoing $\tilde F_4$ wave
decays exponentially on approach to the black hole.

\subsection{Time reflection properties of the Teukolsky equation}

The different forms of Eq's. (\ref{teukopsi0s0}) and (\ref{teukopsi4s4}), or
Eq's. (\ref{eq:tildF_0}) and (\ref{eq:tildF_4}), make it clear that the
Teukolsky equation for the Weyl component $\psi_0$ in the outgoing null
direction $l^a$ and the Teukolsky equation for the Weyl component $\psi_4$ in
the ingoing null direction $n^a$ are not related in a way which makes manifest
the time reflection symmetry ${\cal T}$ of the background Schwarzschild
geometry, defined by ${\cal T}(t,r,\theta,\phi)=(-t,r,\theta,\phi)$ in
Schwarzschild coordinates or by ${\cal T}(\tilde u,r,\theta,\phi)=(-\tilde
v,r,\theta,\phi)$ in terms of Bondi retarded and advanced times $\tilde u
=t-r^*$  and $\tilde v =t+r^*$. The time reflection symmetry could be
incorporated explicitly by introducing null tetrad vectors $L^a=\alpha l^a$ and
$N^a=(1/\alpha) n^a$ satisfying ${\cal T}L^a=-N^a$. However, the explicit form
of the required boost,
\begin{equation}
      \alpha = -\frac {2M}{u} \sqrt{ \frac{2(r-2M)}{r}  }
             = -\frac {2M}{u} \sqrt{ \frac{-2\lambda u}{8M^2-\lambda u}  } ,
\end{equation}
makes it clear that such a time symmetric formulation would introduce
singular behavior at both the black and white hole horizons.

However, this time symmetric tetrad is useful for formulating the time
reflection symmetry of solutions of the Teukolsky equations using other
tetrads. Let $\Psi_4=C_{abcd}N^am^bN^c m^d = \Psi(\tilde u,r,\theta,\phi)$  be
a perturbative solution for $\Psi_4$. Then the time reflection symmetry implies
that $\Psi_0=C_{abcd}L^a\bar m^bL^c\bar m^d = \bar \Psi(-\tilde
v,r,\theta,\phi)$ is a perturbative solution for $\Psi_0$. This correspondence
maps a retarded solution (no incoming radiation) for $\Psi_4$ into an advanced
solution (no outgoing radiation) for $\Psi_0$.  In terms of the $l^a$ and $n^a$
Weyl components, the solution $\psi_4=\psi(\tilde u,r,\theta,\phi)$ corresponds
under time reflection to the solution
\begin{equation}
   \psi_0=\frac {r^2}{256 M^2}e^{r/m}\bar \psi(-\tilde v,r,\theta,\phi).
\label{eq:trefl}
\end{equation}

\section{Null data for a Schwarzschild perturbation}
\label{sec:data}

The Weyl component $\psi_0$ can be posed as constraint-free gravitational
data on an outgoing null hypersurface. Similarly, $\psi_4$  constitutes
constraint-free gravitational data on an ingoing null hypersurface.
These {\em nonlinear} results extend to linearized theory but care must
be exercised in applying them to the Teukolsky equations.

In the Cauchy formulation, the Teukolsky equation for $\psi_4$ is normally
chosen in order to investigate the outgoing radiation introduced by a
perturbation.  However, the Hamiltonian and momentum constraints prevent the
free specification of $\psi_4$ (or  $\psi_0$) on a Cauchy hypersurface.
Consistent Cauchy data for $\psi_4$ must be provided indirectly  by a 3-metric
and extrinsic curvature that solve the constraints. In the Cauchy approach to
the close approximation, this has been provided by (a limit of) Misner's time
symmetric wormhole data~\cite{Misner60}.

In the double-null formulation of the characteristic initial value problem,
data are given on a pair of intersecting null hypersurfaces, one outgoing and
one ingoing. Null data for the Teukolsky equation for $\psi_4$ can be freely
posed on the ingoing null hypersurface but data for $\psi_4$ on the outgoing
null hypersurface has to obtained indirectly. This can be done, as in the
Cauchy problem, by first considering consistent metric data in double null
coordinate, from which the Weyl data for $\psi_4$ can be constructed on both
hypersurfaces. This is the method we use here to generate two examples of
double-null data for the Teukolsky equation:  Robinson-Trautman perturbations
and close approximation data.

\subsection{Robinson-Trautman perturbations}

The Robinson-Trautman space-times~\cite{RT} describe an algebraically special
but distorted and radiating black hole. They provide an important testbed for
the computation of a general perturbative solution by numerical evolution. In
the case of outgoing radiation from a black hole of mass $M$, the metric can
be put in the Bondi form~\cite{robjef}
\begin{equation}
  ds^2 = -({\cal K}-{2M\over r{\cal W}})d\tilde{u}^2-2{\cal W}d\tilde{u}dr
         -2r{\cal W}_{,A}d \tilde{u}dx^A+r^2q_{AB}dx^A dx^B,
\label{eq:rtout}
\end{equation}
where ${\cal K}={\cal W}^2[1-L^2(\log {\cal
W})]$, $L^2$ is the angular momentum operator and ${\cal W}(\tilde{u},x^A)$
satisfies the nonlinear equation
\begin{equation}
      12M \partial_{\tilde{u}}(\log {\cal W}) = {\cal W}^2 L^2 {\cal K}.
    \label{eq:rteq}
\end{equation}

The outgoing Eddington-Finkelstein form of the Schwarzschild metric
Eq.~(\ref{eq:outEF}) results from setting ${\cal W}=1$. More generally, smooth
initial data ${\cal W}(\tilde{u}_0,x^A)$ evolve smoothly to form a
Schwarzschild black hole horizon. The linearized solutions to the
Robinson-Trautman equation (\ref{eq:rteq}) are obtained by setting ${\cal
W}=1+\phi$ and dropping nonlinear terms in $\phi$:
\begin{equation}
    12 M \partial_{\tilde{u}} \phi = L^2(2-L^2)\phi.  \label{eq:rt}
\end{equation}
For a spherical harmonic perturbation $\phi=A(\tilde{u})Y_{\ell m}$ this leads
to the exponential decay $A=A_0 e^{- \tilde{u}
\,\ell(\ell+1)(\ell^2+\ell-2)/12 M}$.

The corresponding Weyl tensor components for the perturbation are
$\psi_0=0$, in agreement with the role of $l^a$ as an algebraically degenerate
principal null direction, and
\begin{equation}
    \psi_4= \frac{2M\,{A_0}\,\left[ \ell(\ell+1)-2  \right] \,
    \left[ -6\,M + \ell\,\left( \ell + 1 \right) \,r \right] \,
    }{3\,r^2 M^2}{\left(-u/M \right) }^
    {-2 + \frac{1}{3} \ell\,\left( \ell + 1 \right) \,
    \left[\ell \left( \ell + 1 \right) -2 \right] }\bar \eth^2 Y_{lm},
\end{equation}
in terms of the affine horizon parameter $u$. The perturbation vanishes on the
black hole horizon ${\cal H}^+$ and is singular at ${\cal I}^-$. This supplies
the
data on ${\cal H}^-$ and an outgoing null hypersurface $u=u_-$ for the
evolution of $\psi_4$ forward in retarded time.

For the corresponding time reversed solution,
$\psi_4=0$. By applying to the Robinson-Trautman perturbations a procedure for
mapping an outgoing solution of Einstein's equations into an ingoing
version~\cite{excise}, we find the solutions
\begin{equation}
  ds^2 = -({\cal L}-{2M\over r{\cal V}})d\tilde{v}^2+2{\cal V}d\tilde{v}dr
         +2r{\cal V}_{,A}d \tilde{v}dx^A+r^2q_{AB}dx^A dx^B,
\label{eq:rtin}
\end{equation}
where $\tilde v$ is the advanced Bondi time coordinate, ${\cal L}={\cal
V}^2[1-L^2(\log {\cal V})]$ and
\begin{equation}
      12M \partial_{\tilde{v}}(\log {\cal V}) =- {\cal V}^2 L^2 {\cal L}.
    \label{eq:rteq2}
\end{equation}

The linearized solutions
obtained by setting ${\cal V}=1+\phi$ satisfy
\begin{equation}
    12 M \partial_{\tilde{v}} \phi = -L^2(2-L^2)\phi.  \label{eq:rt2}
\end{equation}
For a spherical harmonic perturbation $\phi =B(\tilde{v})Y_{\ell m}$, this
leads to
the exponential growth $B=B_0 e^{\tilde{v} \,\ell(\ell+1)(\ell^2+\ell-2)/12
M}$.
The corresponding perturbative Weyl tensor component is
\begin{equation}
  \psi_0 =\frac{B_0\,e^{r/M}\,[\ell (\ell+1)-2 ] \,
    [( -6\,M + \ell\,\left( \ell +1 \right) \,r ] \,
    {(v/M)}^{-2 + \frac{1}{3}\ell\,\left( \ell + 1  \right) \,
     \left[ \ell (\ell+1)-2  \right] }}{384\,M^3}\eth^2 Y_{lm},
\end{equation}
in terms of $v = M e^{\tilde v/4M}$ (the affine parameter along the black hole
horizon). This perturbation vanishes on the white hole horizon and is singular
at
${\cal I}^+$. Nevertheless, it can be used to check a (forward in retarded
time)
evolution algorithm, beginning at a retarded time $u_-$,  by pasting
asymptotically
flat initial null data outside some radius to interior Robinson-Trautman data.

\subsection{Close limit initial data}
\label{sec:closedat}

The coordinates introduced by Sachs to formulate the double-null characteristic
initial value problem~\cite{sachsdn} are especially useful when one
of the null hypersurfaces is a white hole horizon ${\cal H}^-$. Sachs'
coordinate system consists of (i) an affine null coordinate $u$ along the
generators of ${\cal H}^-$, which foliates ${\cal H}^-$ into cross-sections and
labels the corresponding outgoing null hypersurfaces emanating
from the foliation; (ii) angular coordinates $x^A$ which are constant both
along the generators of ${\cal H}^-$ and along the outgoing rays and (iii) an
affine parameter $\lambda$ along the outgoing rays normalized by
$\nabla^{\alpha}u \nabla_{\alpha}\lambda =-1$, with $\lambda =0$ on ${\cal
H}^-$. In the resulting $x^{\alpha}=(u,\lambda ,x^A)$ coordinates, the metric
takes the form
\begin{equation}
   ds^2  = -(W -g_{AB}W^A W^B)du^2
       -2dud\lambda -2g_{AB}W^Bdudx^A +  g_{AB}dx^Adx^B.
\label{eq:smet}
\end{equation}
In addition, it is useful to
set $g_{AB}=r^2h_{AB}$, where $\det(h_{AB})=\det(q_{AB})$, where
$q_{AB}$ is the unit sphere metric.

The requirement that ${\cal H}^-$ be null implies that $W=0$ on ${\cal H}$ and
the gauge freedom on  ${\cal H}^-$ can be fixed by choosing the shift so that
$\partial_u$ is tangent to the generators, implying that $W^A=0$ on ${\cal
H}^-$, and by choosing the lapse so that $u$ is an affine parameter, implying
that $\partial_\lambda W=0$ on ${\cal H}^-$. In addition to these choices,
we fix the affine freedom in $u$ by specifying it on a slice ${\cal
S}^-$ of ${\cal H}^-$, which is located at an early time approximating the
asymptotic equilibrium of the white hole at past time infinity $I^-$. The
outgoing null hypersurface ${\cal J}^-$ emanating from ${\cal S}^-$ then
approximates past null infinity ${\cal I}^-$. The Schwarzschild metric in
Israel coordinates (\ref{eq:amet}) is obtained in the spherically symmetric
case
when $W^A=0$ and $h_{AB}=q_{AB}$.

The double-null problem for the close limit of a white hole is posed on the
ingoing null hypersurface ${\cal H}^-$ and the outgoing null hypersurface
${\cal J}^-$, which extends to ${\cal I}^+$. In order to pose the double-null
Teukolsky problem for $\psi_4$ in the
perturbative regime, we
generate the data for $\psi_4$ from metric data for
the nonlinear version of the problem. The metric version of the null data
consists of the values of the spin-weight-two field $J=q^A q^B h_{AB}$ on
${\cal H}^-$ and ${\cal J}^-$.
For a perturbation of a Schwarzschild background, $r$ is given by Eq.
(\ref{eq:r}).
On the event horizon ${\cal H}^-$,
$$
   \bar \psi_4 = {\frac{1}{2}} J_{,uu} - {\frac{1}{2}} J_{,u} K K_{,u}
 + {\frac{1}{4}} J K_{,u}^{2} +
  J_{,u} r^{-1} r_{,u} + J r^{-1} r_{,uu} + {\frac{1}{4}} J_{,u}^{2} \bar{J}.
$$
where $K=\sqrt{1+J\bar J}$, which reduces in the linear regime to
\begin{equation}
      \bar \psi_4 \approx {\frac{1}{2}} J_{,uu} .
      \label{eq:jtopsi}
\end{equation}
Off the horizon the expression for $\bar \psi_4$ is more complicated and
involves $W$ and $W^A$ as well as $J$ and $r$.

The horizon data for a head-on fission of a white hole, can be obtained from a
conformal model based upon an ingoing null hypersurface emanating from a
prolate spheroid embedded in a flat space~\cite{ndata}. Let $(\hat r, \theta,
\phi)$ be standard spherical coordinates for the inertial time slices $\hat t
=constant$ of Minkowski space. In the close limit, the eccentricity of the
spheroid vanishes and the Minkowski null hypersurface reduces to the light cone
from a sphere $\hat t=0$, $\hat r= a$. The perturbation of its conformal null
geometry is described, to linear order in the eccentricity, by
\begin{equation}
     J(\hat t, \theta) = -\frac {a \sin^2\theta}{\hat t -a},
\label{eq:closej}
\end{equation}
where the relation between $\hat t$ and the affine
parameter $u$ on the white hole horizon
is
\begin{equation}
  \frac{d\hat t}{d u}= \Lambda (\hat \tau) =\frac {\hat \tau^2 (\hat \tau-1)^2}
      {(3 -5\hat \tau +\hat \tau^2)^2}
             \bigg (\frac {(5-\sqrt{13}) -2\hat \tau}
                 {(5+\sqrt{13})  -2\hat \tau}\bigg )^{4/\sqrt{13}},
\label{eq:affine}
\end{equation}
in terms of
\begin{equation}
   \hat \tau= \frac{\hat t -a}{p}.
\end{equation}
Here $p$ and $a$ are positive parameters which adjust the affine freedom in the
position of the Minkowski null cone on the white hole horizon. At early times
Eq. (\ref{eq:affine}) implies $u\sim \hat t$ but as the Minkowski null cone
pinches off at $\hat t=a$ the corresponding affine time on the white hole
horizon asymptotes to $u\rightarrow \infty$. In terms of the inverted
pair-of-pants picture for a white hole fission, the pants legs are mapped to
$u=\infty$  so that in the close limit the individual white holes are mapped to
$I^+$ along the white hole horizon in the Kruskal manifold. The details are
discussed elsewhere in a treatment of fully nonlinear null data for the general
two black hole problem~\cite{asym,compnull}.

Close limit data for $J(u,\theta)$ on the white hole horizon is determined by
integrating Eq. (\ref{eq:affine}) and substituting into Eq. (\ref{eq:closej}).
These equations allow the rescaling $u\rightarrow Ku$, $\hat t \rightarrow K
\hat t$, $p\rightarrow K p$ and $a\rightarrow K a$ which allow us to set $p=1$
without any loss of generality. Note, that the rescaling $u\rightarrow K u$ is
equivalent to the time translation isometry $\tilde u \rightarrow  \tilde u +
const$. In order to eliminate nonessential parameters, we initiate the
integration at the bifurcation sphere $u=0$. Then, up to scale, the close data
are determined by $\hat \tau_0 = \hat \tau |_{u=0} <0$ or in terms of $J$ by
the parameter
\begin{equation}
   \eta = -\frac {p J|_{u=0} }{ u J|_{u=\infty} }= -\frac {1}{\hat \tau_0},
   \label{eq:eta}
\end{equation}
which is independent of the overall scale freedom $J \rightarrow \lambda J$
that is factored out in the close approximation. The parameter $\eta$ is a
scale invariant parameter describing the physical properties of the close
limit. It determines the yield of the white hole fission. In the time reversed
scenario of a black hole collision, $\eta$ would be related to the inelasticity
of the collision. No similar parameter seems to appear in the Cauchy
description of the close approximation in terms of time symmetric Misner
data~\cite{pp}.

The close limit data on the horizon for $\psi_4$ used in the simulations
presented in Sec. \ref{sec:clwave} are obtained by integrating Eq.
(\ref{eq:affine}) with a 4th order Runga-Kutta scheme and carrying out the
substitutions into Eq's (\ref{eq:closej}) and (\ref{eq:jtopsi}). The data on an
early outgoing null hypersurface are accurately approximated by setting $\psi_4
=0$ since Eq. (\ref{eq:closej}) implies $\psi_4 = O(u^{-3})$. This approximates
the condition on the data that there be no ingoing radiation at ${\cal I}^-$.
\section{Numerical algorithm}
\label{sec:algorithm}

Before giving the details of the numerical algorithm, we should state the goals
we want to achieve. For many purposes, it would seem sufficient to evolve the
waveform until one  can read off the first few cycles of quasinormal mode
oscillation. This is sufficient in practice to compare with a nonlinear
evolution, to get the astrophysically relevant part of the waveform, to compare
quasinormal mode results with those in the literature, etc. Instead we define
as our criterion of quality the ability to evolve stably and accurately well
into the domain where the waveform is dominated by a power law, which requires
at least a $1000M$ of Bondi time for our typical data. This turns out to be a
rather stringent criterion, which rules out a number of numerical approaches
which we have tried. In all such approaches, our overall strategy is to
compactify the outgoing null direction and bring ${\cal I}^+$ into a finite
coordinate distance while maintaining regularity of the equations.

We begin the description of our numerical setup with a discussion of the
ingoing null geodesics, which forms the basis  of  our approach to the
numerical solution of the Teukolsky equation. Then we briefly describe a few of
the algorithms which do not work completely satisfactorily for the  Teukolsky
equation, and explain why this is so. We believe that this also provides useful
experience for nonlinear studies, where, lacking a stationary background
geometry, the source of problems may be much less obvious.

Our present results pertain to the Teukolsky equation for $\psi_4$, where
$\psi_4$ describes the outgoing radiation through its asymptotic $O(1/r)$
behavior at ${\cal I}^+$. The incoming radiation at ${\cal I}^-$ is described
by $\psi_0$, which has asymptotic $O(1/r^5)$ behavior at ${\cal I}^+$.  As a
result, an accurate treatment of the Teukolsky equation for $\psi_0$ requires
different numerical methods, which will be described in a forthcoming paper.

\subsection{Ingoing null geodesics}

Null geodesics are fundamental to the design of the numerical algorithm since
they are the characteristics of the Teukolsky equation. In particular, since we
handle the angular part of the spin-weight-zero Teukolsky equation by a
spherical harmonic decomposition in which $\eth \bar \eth = -\ell(\ell+1)$, the
relevant characteristics are the radial null geodesics. The outgoing null
geodesics are automatically built into the characteristic evolution scheme,
which is based upon a retarded time foliation. The remaining issue is how to
effectively incorporate the behavior of the ingoing null geodesics into the
algorithm.

Consider first the description of the ingoing null geodesics in a compactified
version of Israel coordinates $(u,x)$, where $x=\lambda/(M+\lambda)$ so that
${\cal I}^+$ is located at $x=1$.  The analytical simplicity of these
coordinates is not matched by their numerical convenience. The ingoing null
geodesics satisfy
$$
   \frac{d x}{du} = -(1-x)^2 \frac{W}{2 M}
                 =
      \frac{(1-x) x^2}{8 M(1-x)-u x}.
$$
Near ${\cal I}^+$ where $1-x =\delta <<1$, this reduces to
\begin{equation}
   \label{eq:geodesics_asymptotics}
   \frac{d \delta}{d u} = \frac{\delta}{u}.
\end{equation}
Hence $\delta$, and the separation between neighboring geodesics near $\cal
I^+$, decays linearly with $u$ and exponentially with $\tilde u$. Thus
evolving for a Bondi time of $\tilde u = 1000M$ is impossible as the geodesics
would be within $e^{-250} \delta_0$ of each other.

Not only is the $x$-coordinate numerically unsatisfactory in the way it
compactifies $\cal I^+$, the $u$-coordinate is also inconvenient in the way it
covers the exterior Kruskal quadrant in a finite retarded time. This prevents
the long term numerical resolution of the ringdown (with the characteristic
time scale of the lowest quasinormal mode) without using an exponentially
decaying step size $\Delta u$. This is simple to fix by using Bondi time
$\tilde u$ as the time step coordinate.

The problem with the $x$-coordinate can be ``delayed'' by introducing a
dynamical grid, in which the gridpoints move along the ingoing null geodesics,
a strategy that has been successful in studies of spherical critical
collapse~\cite{garf1}. This approach drastically decreases the discretization
error and is sufficient to evolve for about 100 $M$, and read off the
quasinormal ringdown frequency and damping time with good accuracy.  However,
the $\Delta x$ intervals between neighboring geodesics decrease exponentially
and the approach breaks down once the separation between neighboring gridpoints
falls below the error of the geodesic integrator (e.g. machine precision) and a
``numerical crossing'' of the geodesics effectively occurs.

Note that in practice the computation of the null geodesics is related to the
problem of inverting the definition of the tortoise coordinate $$r^* = r + 2 M
\log(\frac{r}{2 M}-1)$$ to compute $r$, which is also required for our
production algorithm as discussed in the next section. Both problems are
handled numerically by solving the above  implicit equation iteratively using
Newton's method (in terms of the appropriate coordinates).

\subsection{Numerical algorithm for outgoing radiation}

The preceding considerations lead us to the following choice of algorithm for a
production level unigrid code with optimal performance. It is based on a
$(\tilde u,\rho)$ coordinate system, where $\rho$ is a radial coordinate, which
compactifies ${\cal I}^+$, defined implicitly by
\begin{equation}
       r^*=\rho_0 \tan \rho,
\end{equation}
with $\rho_0$ an adjustable parameter and $-\pi/2 \leq \rho \leq
\pi/2$.  The $\rho$ coordinate allows good resolution at all times
near both ${\cal I}^+$ and the white hole horizon, .

In this coordinate system, we evolve the $\ell$th spherical harmonic component
of $\tilde F_4$ by expressing the
second order differential equation (\ref{eq:tildF_4})
as the two coupled equations
\begin{eqnarray}
                      \partial_\rho\tilde F_4  &=& G
		      \label{eq:F4} \\
	\partial_{\tilde u} G
	    &=& \frac {\cos^2 \rho}{2 \rho_0} \partial_\rho G +
           \left ( 2 \frac{ (r-3 M)}{r^2}
	   - \frac{\sin \rho \cos \rho}{\rho_0} \right) G -
             \frac{\rho_0}{\cos^2 \rho} \frac{(r-2 M)
	     ((\ell^2 +\ell-2) r +6 M )}{2 r^4} \tilde F_4 .
	     \label{eq:Gdot}
\end{eqnarray}
The background mass $M$ can be scaled out of the above equations by the
rescaling $\rho_0 \to M \rho_0$, $\tilde u \to M \tilde u$ and $r \to M r$. In
this way, simulations can be carried out with $M=1$ without loss of generality.
Note that rescaling $\tilde u$ is independent of the rescaling $u\rightarrow
Ku$ (see Sec. \ref{sec:closedat}) which generates the translation isometry of
$\tilde u$.

The relatively sensitive features of Eq. (\ref{eq:Gdot}) on a
hypersurface of constant $\tilde u$ are located in the region near
$r=2M$. This is the  main reason why the $\rho$ coordinate is so useful
since it concentrates grid points in that region while maintaining a
uniform grid spacing. We make the choice $\rho_0 =40 M$, which gives
good resolution throughout the evolution, well past the ringdown phase
and into the final power law tail.     Note that this would not be
possible with the simpler approach of writing the Teukolsky equation in
(a compactified version of) double null coordinates $(\tilde u$,
$\tilde v)$, which gives excellent results until one reaches the power
law tail. At this stage the dynamics is essentially dominated by the
``Schwarzschild potential'', which is not well resolved in double null
coordinates.

Equation (\ref{eq:F4}) is solved using a second order accurate
integration in $\rho$. Equation (\ref{eq:Gdot}) is solved using a second order
(in time) Runga-Kutta scheme, with the $\partial_\rho G$ term evaluated by
means of second order accurate forward differencing in the interior of the
grid, and second order accurate central differencing for the point neighboring
$\cal I^+$. (The $\partial_\rho G$ term drops out on both $\cal H^-$ and $\cal
I^+$, where $\partial_{\tilde u}$ is a characteristic direction).

For a typical choice of initial data  the power law  tail only sets in after
the quasinormal oscillations have decayed by more than 10 orders of magnitude.
In order for the final tail not to be lost in machine error it is necessary to
evolve the quasinormal phase in quadruple  precision.

\section{Waveforms}
\label{sec:waveforms}

In this section, we present computed waveforms for three types of quadrupole
data, with the background mass scaled to $M=1$. The first case, an analytically
known Robinson-Trautman perturbation, is used to establish second order
convergence of the numerical algorithm. The Robinson-Trautman waveform decays
as a pure exponential. The second case, a pulse of compact support emanating
from the white hole horizon, serves to monitor the ability of the code to track
many cycles of the quasinormal ringdown of a generic radiation tail. The final
case is the close limit waveform from a white hole fission.

\subsection {Robinson-Trautman testbed}

\begin{figure}
\centerline{\epsfxsize=6in\epsfbox{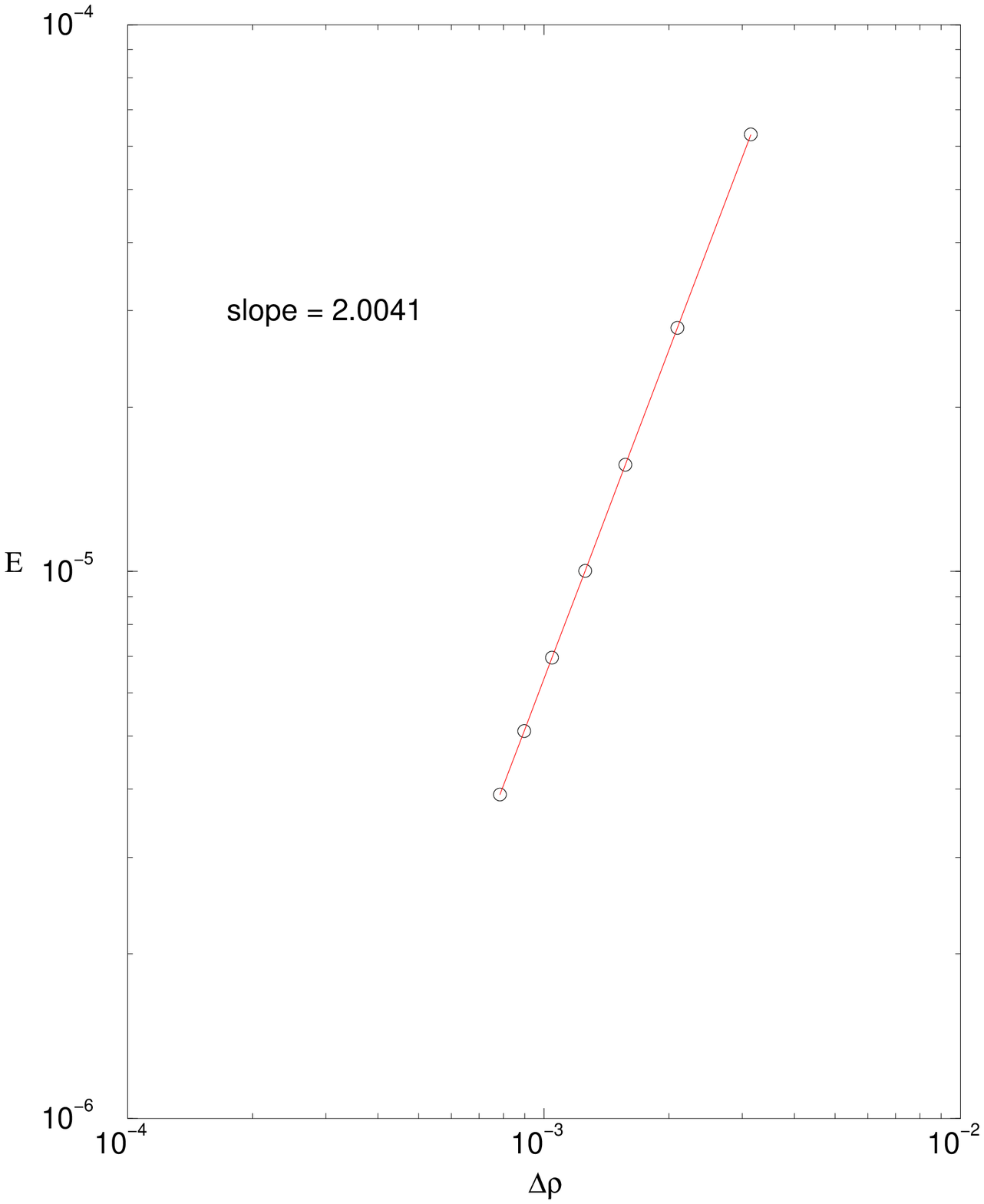}}
\caption{Robinson-Trautman convergence test: $\ell_\infty$ error norm
$E$ versus gridsize $\Delta \rho$.}
\label{F_4_RT_L}
\end{figure}

An $\ell=2$ Robinson-Trautman perturbation is determined by
the spin-weight-0 field
\begin{equation}
   \tilde F_4 =\frac {r-1}{r} e^{-2 \tilde u} .
\end{equation}
We use this to perform a convergence test of the code by evolving from  $\tilde
u =0$ to $\tilde u =5$ and then examining the $\ell_{\infty}$ norm  of the
error $E=||\tilde F_{4,NUMERIC}-\tilde F_{4,ANALYTIC}||$  versus grid
spacing at $\tilde u =5$, at which time the signal has decreased by a factor of
$~ e^{-10}$. The convergence plot of the error given in Fig.~\ref{F_4_RT_L}
determines a slope of $2.0041$, in excellent agreement with the theoretical
second order accuracy of the algorithm.

\subsection{Ringdown from a compact pulse}
\label{sec:compulse}

\begin{figure}
\centerline{\epsfxsize=6in\epsfbox{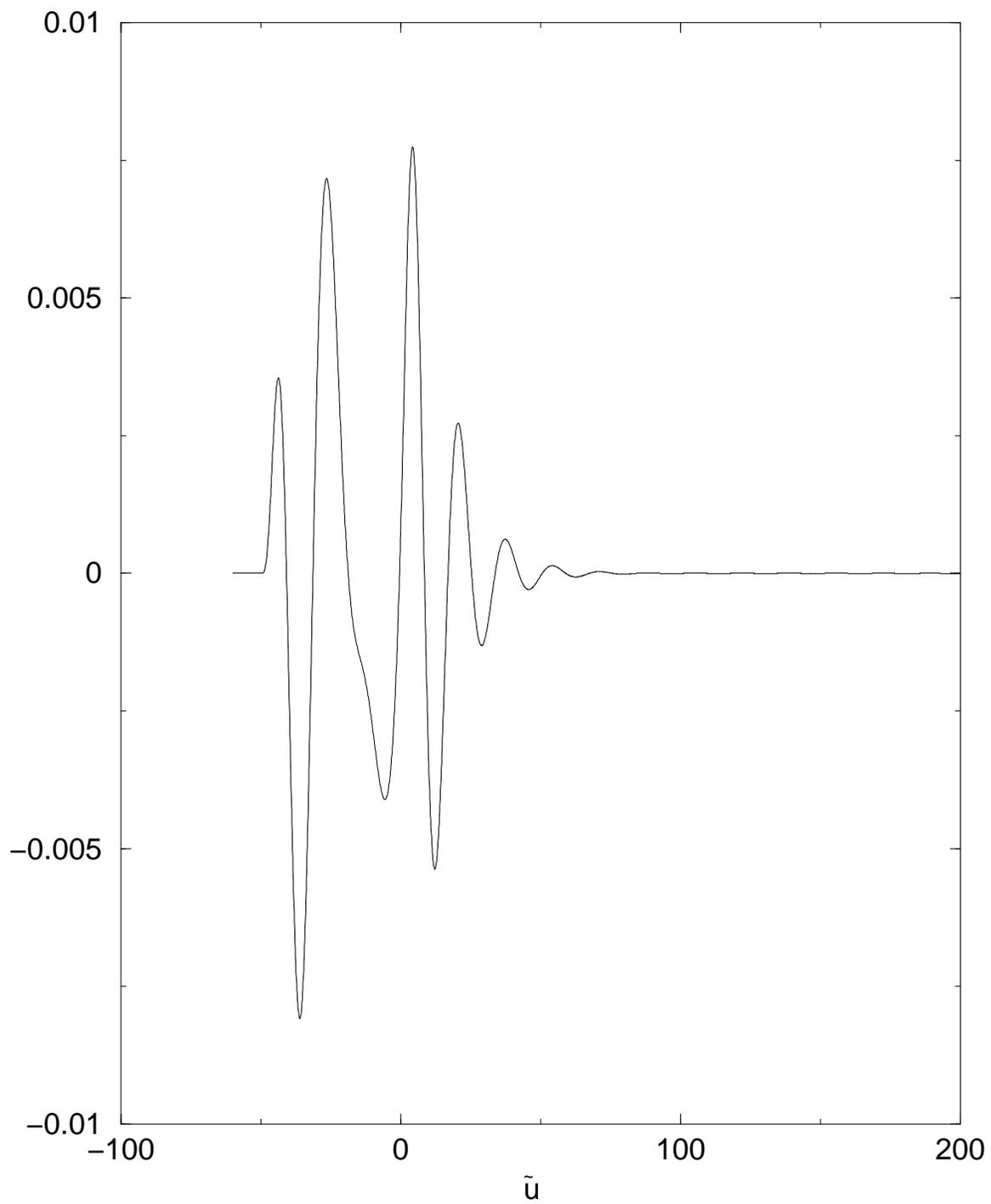}}
\caption{Waveform $\tilde F_4(\tilde u)$ at ${\cal I}^+$ produced by a
single pulse emerging from the white hole horizon.}
\label{F_4_wave}
\end{figure}

We simulate the evolution of an $\ell =2$ quadrupole pulse of compact support
emerging from the white
hole horizon $\cal H^-$.
The pulse consists of a single peak of the form
\begin{equation}
    \tilde F_4(\tilde u, x=0) =
        ((\tilde u -\tilde u_{min})(\tilde u_{max} - \tilde u))^4 \, ,
	\,\,\,\,\,  \tilde u_{min} < \tilde u < \tilde u_{max},
\end{equation}
with $\tilde F_4(\tilde u, x=0 )=0$ outside this interval. Figure
\ref{F_4_wave} shows the waveform on ${\cal I}^+$ obtained by evolving $\tilde
F_4$ with initial data $\tilde F_4=0$  on an outgoing hypersurface preceding
the pulse. In the simulation used to produce the waveform at ${\cal I}^+$, we
choose $\tilde u_{min}=-50$ and $\tilde u_{max}=0$ and evolve from $\tilde
u=-60M$ to $\tilde u=2000 M$. The simulation was performed in two steps. The
first step, in the interval from $\tilde u = -60 M$ until $\tilde u = 250 M$,
was
performed in quadruple precision in order not to lose the final tail in
roundoff error. The second step, in the interval from $\tilde u =250 M$ until
$\tilde u = 2000 M$, was performed in double precision.

\begin{figure}
\centerline{\epsfxsize=6in\epsfbox{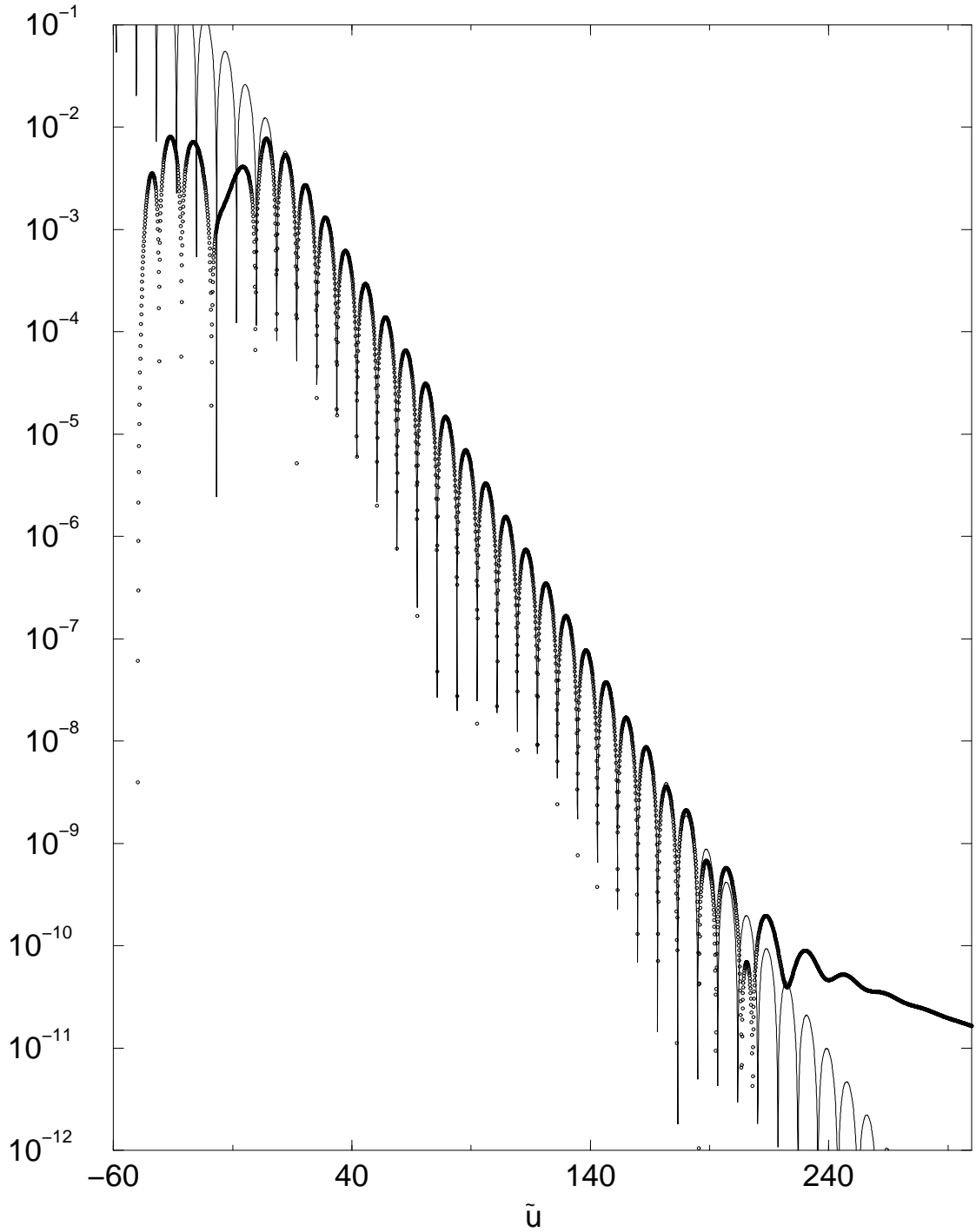}}
\caption{Log plot of $|\tilde F_4(\tilde u)|$ for the waveform in Fig.
\ref{F_4_wave} (darker curve) and a fit to the quasinormal ringdown (lighter
curve).}
\label{F_4_poly_log}
\end{figure}

Figure \ref{F_4_poly_log} is a logarithmic plot of the absolute value of the
waveform versus $\tilde u$ for the same data. It covers the period from the
onset of quasinormal ringdown to the onset of the final tail decay. The
logarithmic plot clearly demonstrates the exponential decay and shows a fit to
a  quasinormal decay. The lowest quasinormal mode for a gravitational
perturbation of the Schwarzschild metric has the theoretical form $ f \sim
\sin(.373672 \tilde u) \exp(-.0889625 \tilde  u)$~\cite{schmidtnoll}. The fit
of the computed waveform to a quasinormal decay is $\tilde F_4 \sim
\sin(.373668  \tilde u)\exp(-.088951 \tilde u)$, in excellent agreement with
the expected theoretical form. The corresponding fit of the close approximation
waveform given in Sec. \ref{sec:clwave} yields the quasinormal dependence
$\tilde F_4 \sim  \sin(.3736735 \tilde u)\exp(-.0889575  \tilde u)$. A
conservative comparison of these two  calculations indicates a quasinormal
dependence  $\tilde F_4 \sim  \sin(.37367  \tilde u)\exp(-.08895  \tilde u)$,
with the numerical uncertainty in the last digit.

\begin{figure}
\centerline{\epsfxsize=6in\epsfbox{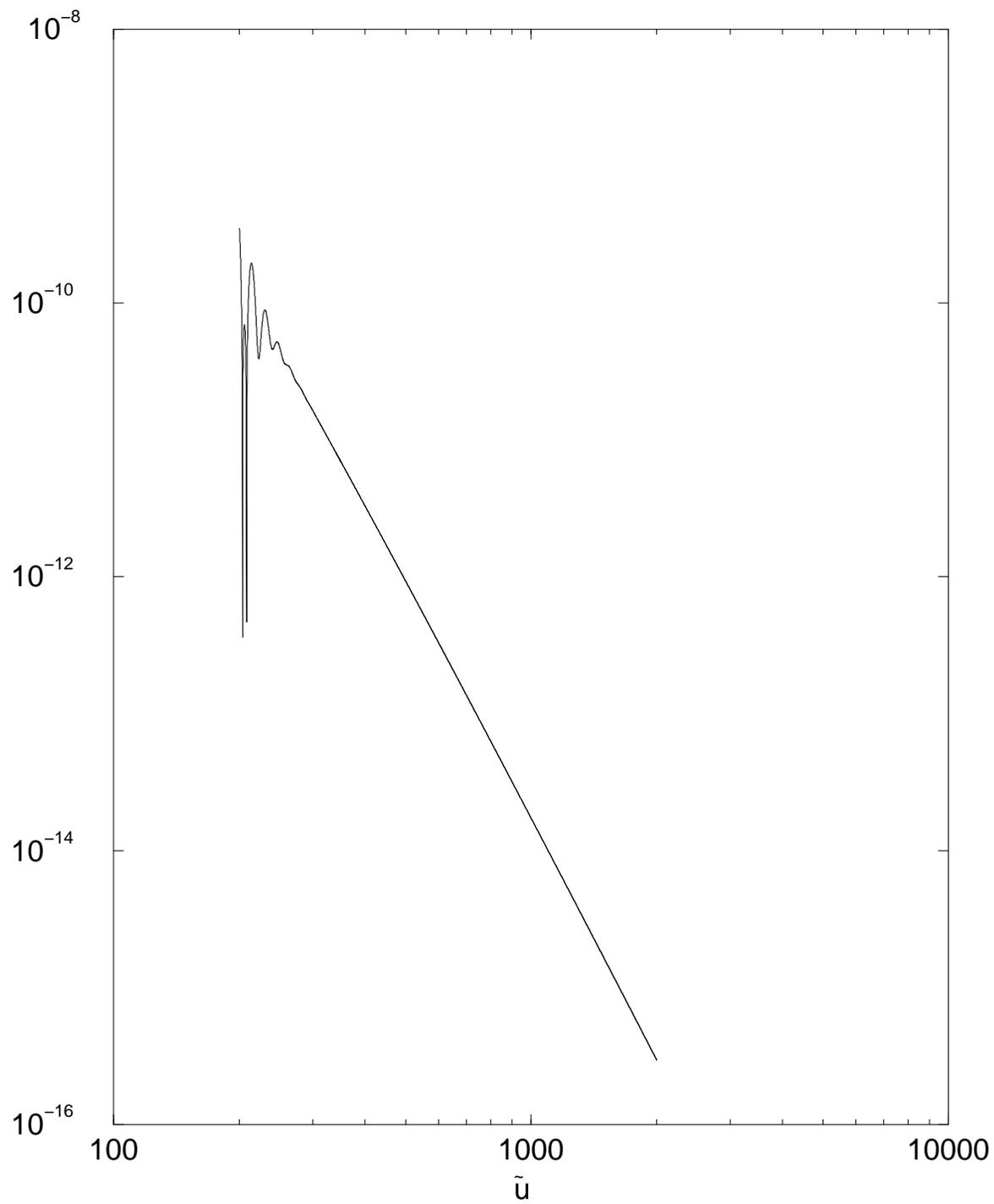}}
\caption{Log-log plot of the tail of the waveform in Fig. \ref{F_4_wave}.
}
\label{F_4_poly_log_log}
\end{figure}

Figure \ref{F_4_poly_log_log} shows a log-log plot of the  late time behavior
of the waveform and the final tail. The measured slope of the  tail indicates a
power law decay ,  with the power varying from $\tilde F_4 \propto \tilde u
^{-5.76}$ near the beginning of  the tail to $\tilde F_4 \propto \tilde u
^{-5.89}$ near the end.

\subsection{The close approximation waveform}
\label{sec:clwave}

As discussed in Sec. \ref{sec:closedat}, the effective parameter space for the
head-on close approximation data can be reduced to the single scale invariant
parameter $\eta$ controlling the fission yield. In the simulations presented
here we set the scale dependent parameter $p=1$. Thus, in accord with
the discussion in Sec. \ref{sec:closedat}, we identify $\hat t -a=\hat \tau$
and prescribe the horizon data used in the simulations
in the normalized form
\begin{equation}
  F_4 = -(\Lambda \partial_{\hat \tau})(\Lambda \partial_{\hat \tau})
            \frac{1}{\hat \tau},
\label{eq:closej2}
\end{equation}
after factoring out the $\ell=2$ angular dependence.

The time dependence of the close approximation data is quite mild when
expressed as a function of $\hat \tau$, as in Eq.~(\ref{eq:closej2}). However,
the relationship (\ref{eq:affine}) can cause the dependence on $u$ to be quite
sharp. There is a transition region where the behavior of $\hat \tau (u)$
changes from the asymptotic form  $d\hat \tau /du  \to 1$ as $\hat \tau \to -
\infty$ to $d\hat \tau /du  \to 0$ as $\hat \tau \to 0$. For large values of
the parameter $\eta$, this produces sharply pulse shaped data, as described
below.

\begin{figure}
\centerline{\epsfxsize=6in\epsfbox{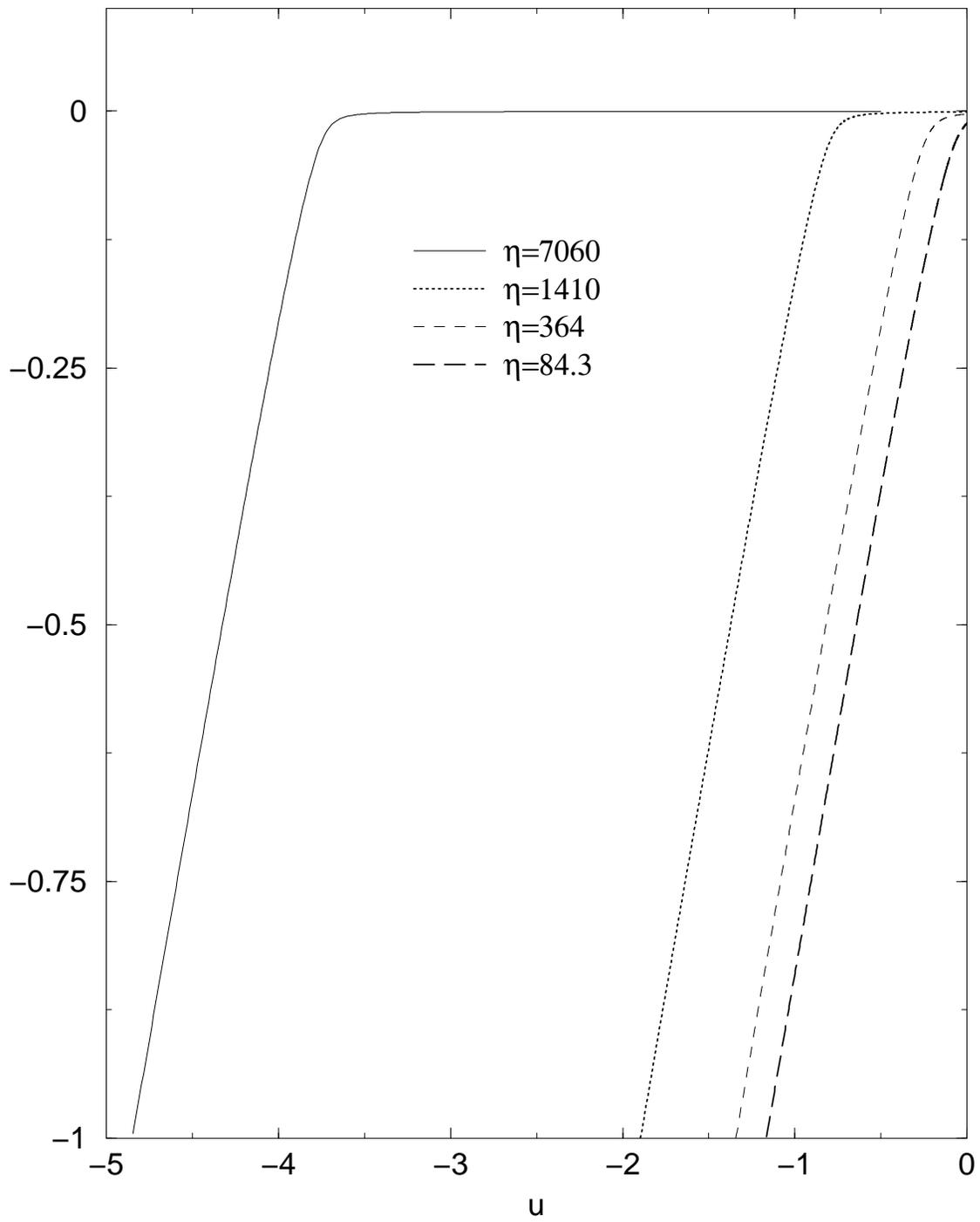}}
\caption{$\hat \tau (u)$ versus $u$ for 4 values of  $\eta$. }
\label{tau_versus_eta}
\end{figure}

Figure \ref{tau_versus_eta} plots $\hat \tau$ versus $u$ for $\eta= 7060$,
$1410$, $364$, and $84.3$. The plots reveal a relatively sharp transition in
the slope. This transition region is translated in the negative
$u$-direction as $\eta$ increases. For sufficiently small $\eta$ the
transition occurs at $u>0$, in the region of the white hole horizon which does
not affect the exterior spacetime.

\begin{figure}
\centerline{\epsfxsize=6in\epsfbox{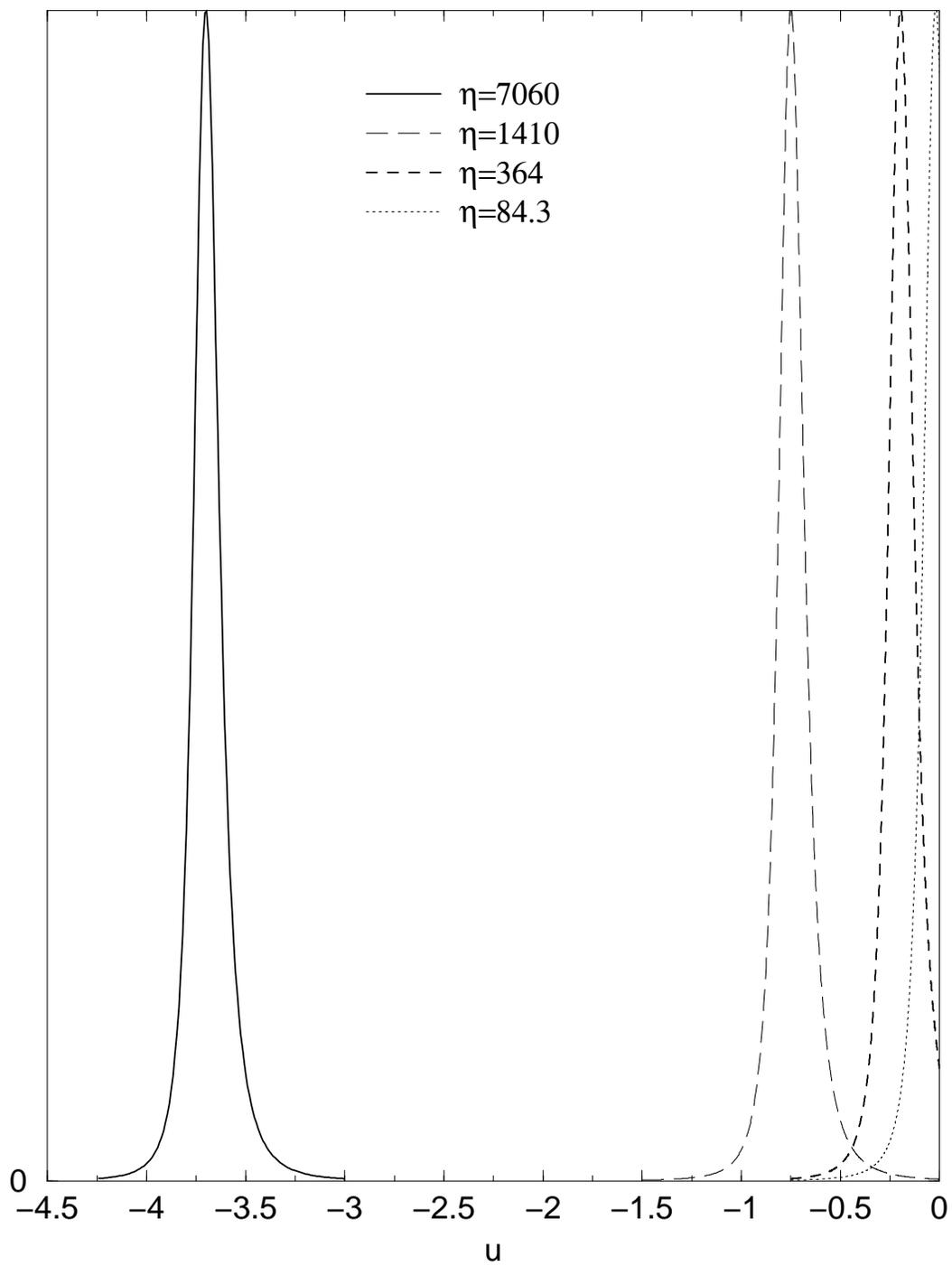}}
\caption{Close approximation data $F_4(u)$ on $\cal H^-$ for 4 values of
$\eta$.}
\label{psi_4_u_versus_eta}
\end{figure}

The location of the transition region affects the nature of the horizon data.
Figure \ref{psi_4_u_versus_eta} shows $F_4(u)|_{\cal H^-}$ for the
above  values of $\eta$. The value of $\eta$ only changes the position of the
transition region, not its width. Hence a change in $\eta$ translates the
horizon data $F_4(u)|_{\cal H^-}$ but does not change its shape.

\begin{figure}
\centerline{\epsfxsize=6in\epsfbox{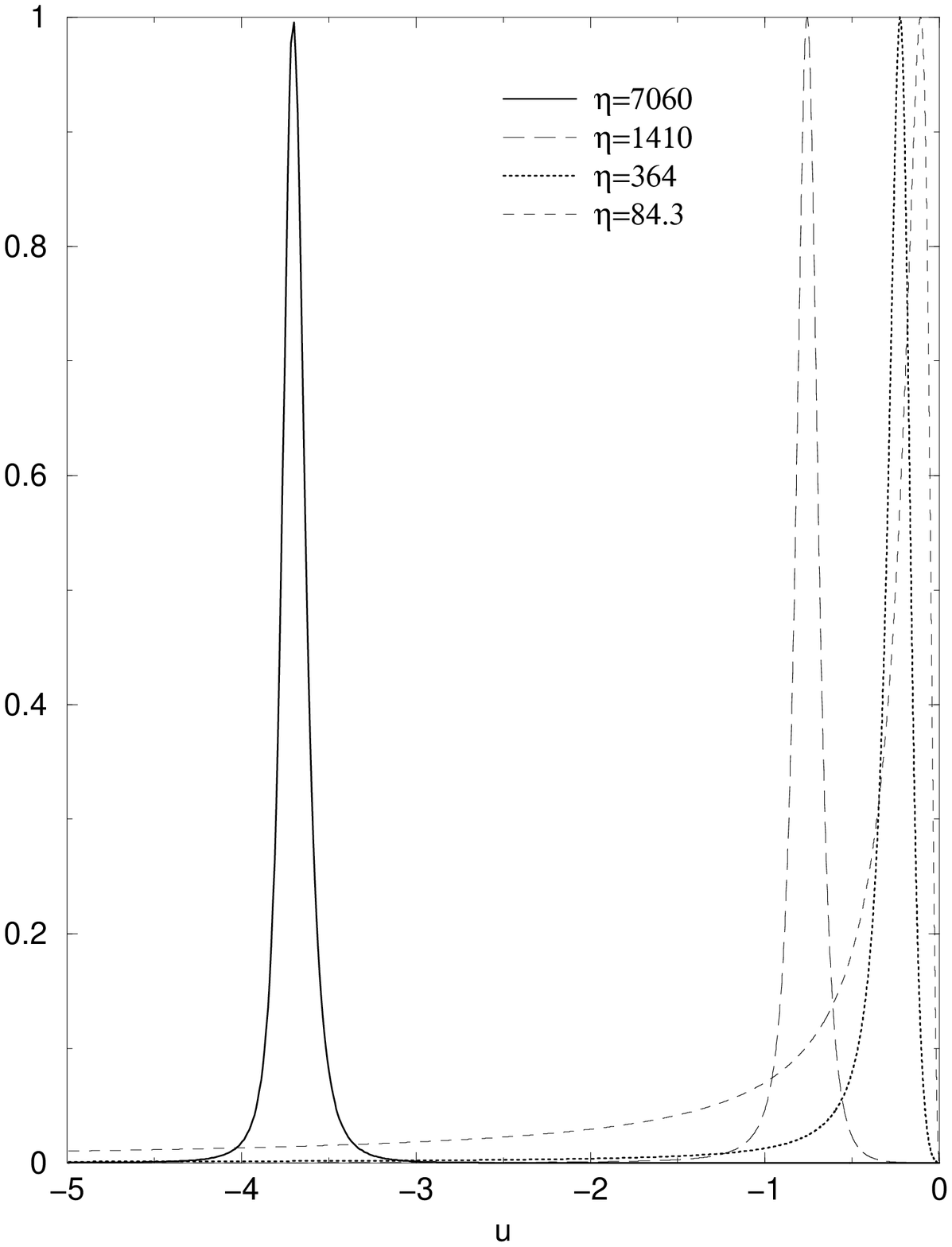}}
\caption{Close approximation data $\tilde F_4(u)$ on $\cal H^-$ for
$\eta=$ 7060, 1410,  364 and 84.3, with the amplitudes renormalized by the
relative factors of 1, 24.03,
305.9 and 3223, respectively.}
\label{tilde_psi_4_u_versus_eta}
\end{figure}

\begin{figure}
\centerline{\epsfxsize=6in\epsfbox{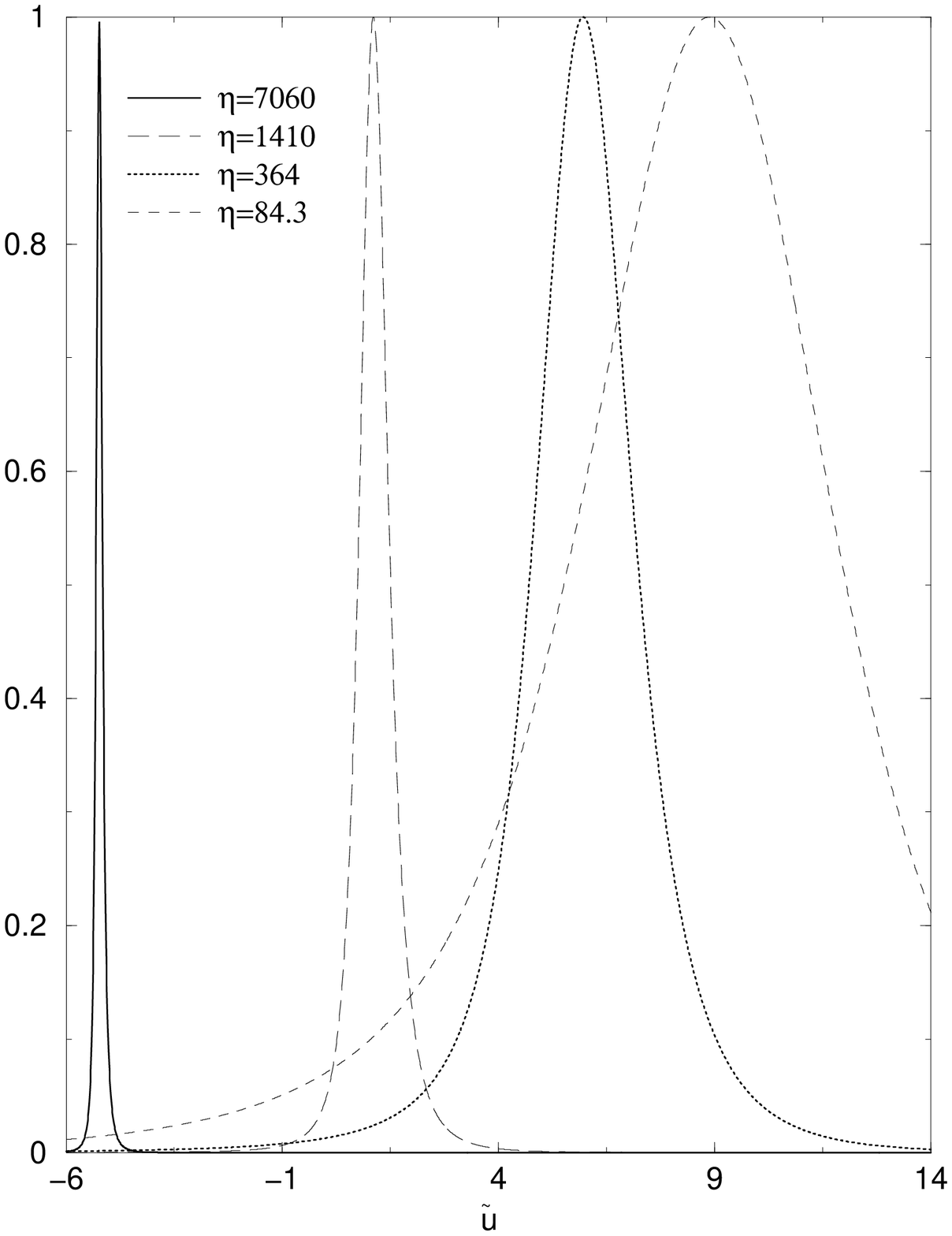}}
\caption{Close approximation data $\tilde F_4(\tilde u)$ on $\cal H^-$ for
$\eta=$ 7060, 1410,  364 and 84.3, with the amplitudes renormalized by the
relative factors of 1, 24.03,
305.9 and 3223, respectively.}
\label{tilde_psi_4_tilde_u_versus_eta}
\end{figure}

\begin{figure}
\centerline{\epsfxsize=6in\epsfbox{
  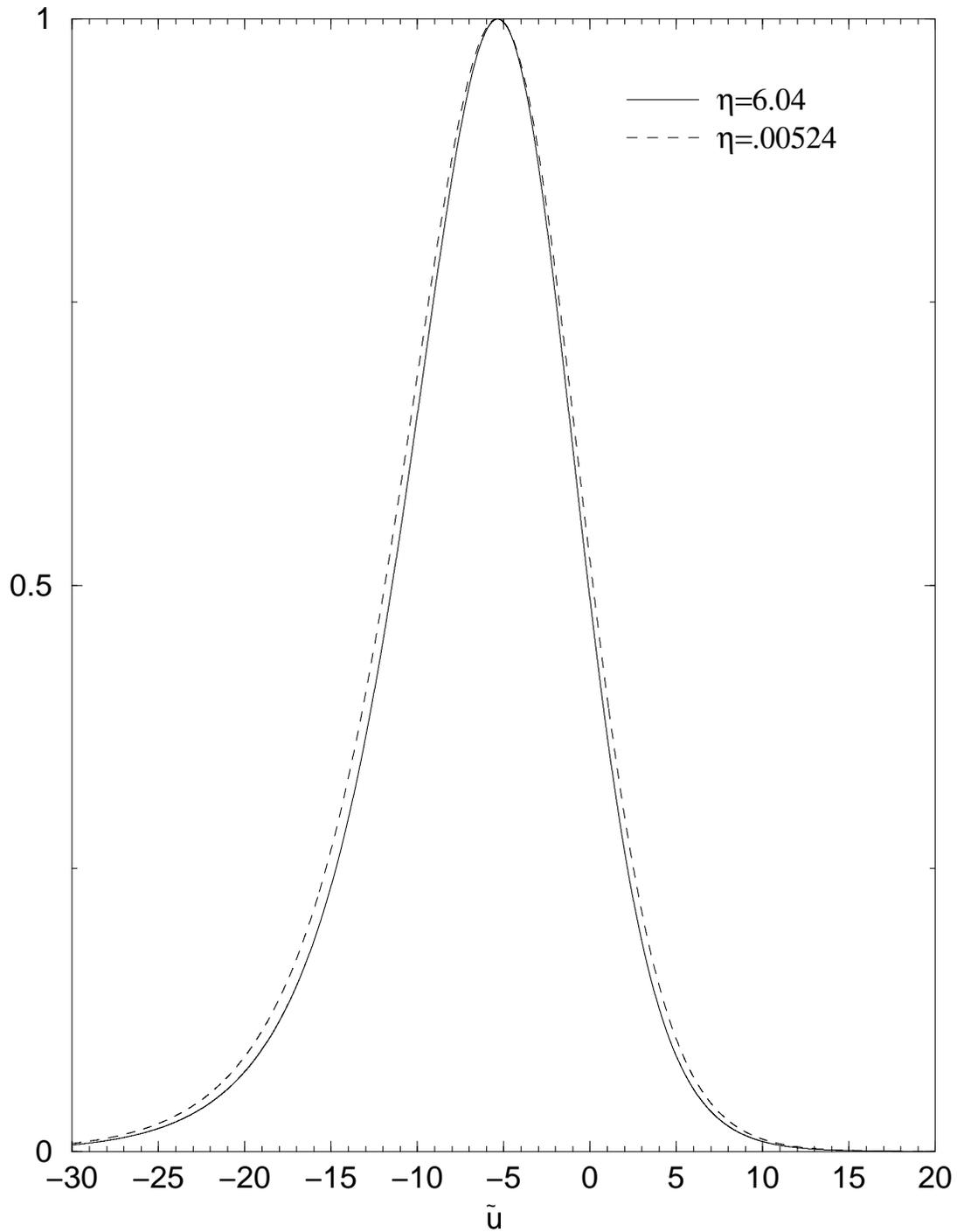}}
\caption{Close approximation data $\tilde F_4(\tilde u)$ on $\cal
H^-$ for $\eta=6.04$ and $.00524$, with  the amplitudes renormalized by the
relative factors of 1 and 1220. }
\label{tilde_psi_4_tilde_u_versus_small_eta}
\end{figure}

The horizon data for $\tilde F_4 (\tilde u)$ has a more complicated dependence
on $\eta$ due to the exponential relationship between $u$ and $\tilde u$ and
the extra factor of $u^2$ introduced by the change in tetrad. This is of
physical importance since it is $\tilde F_4 (\tilde u)$ which is the observed
waveform at ${\cal I}^+$. The factor of $u^2$ suppresses pulses centered at
smaller $|u|$ compared to those centered at larger $|u|$ and forces the
resulting pulse to vanish at $u=0$. The relation between $\tilde u$ and $u$
varies from an exponentially increasing blueshift at large negative $u$ to an
exponentially increasing redshift at $u=0$. This has the effect  of
compressing pulses centered at more negative $u$ compared to those centered at
less negative $u$.  These effects combine to produce successively broader
pulses for successively smaller $\eta$. However, once $\eta$ is sufficiently
small, the transition region is located at $u>0$ and $\eta$ does not affect
the shape of $\tilde F_4 (\tilde u)|_{\cal H^-}$, although it affects its
overall amplitude. In that case $\hat \tau\approx u +\hat \tau_0$ for $-\infty
< u < 0$, and $F_4|_{\cal H^-} \propto 1/(u+\hat \tau_0)^3$, where $\hat
\tau_0 =-1/\eta$.  As a result, modulo a constant overall multiplicative
factor and a constant shift in $\tilde u$,
\begin{equation}
  \tilde F_4(\tilde u)|_{\cal H^-} \propto
      \frac { \eta e^{-\tilde u/2} } { (e^{-\tilde
            u/4} +1)^3}.
	    \label{eq:smalleta}
\end{equation}
Figures \ref{tilde_psi_4_u_versus_eta} and
\ref{tilde_psi_4_tilde_u_versus_eta} show $\tilde F_4(u)|_{\cal H^-}$ and
$\tilde F_4(\tilde u)|_{\cal H^-}$, respectively, for $\eta= 7060$, $1410$,
$364$, and $84.3$.  Figure \ref{tilde_psi_4_tilde_u_versus_small_eta} shows
$\tilde F_4|_{\cal H^-}$ versus $\tilde u$ for small $\eta$, with the
amplitude and position of the peaks adjusted so that they overlap. Except for
the overall  amplitude, there is no significant effect on the data even when
$\eta$ is reduced by 3 orders of magnitude. For small values of $\eta$,
$\tilde F_4(\tilde u)|_{\cal H^-}$ scales linearly with $\eta$ in accord with
Eq. (\ref{eq:smalleta}), whereas for large $\eta$ it scales quadratically, as
evident from the renormalizations in Fig's \ref{tilde_psi_4_u_versus_eta}
and  \ref{tilde_psi_4_tilde_u_versus_eta}.

\begin{figure}
\centerline{\epsfxsize=6in\epsfbox{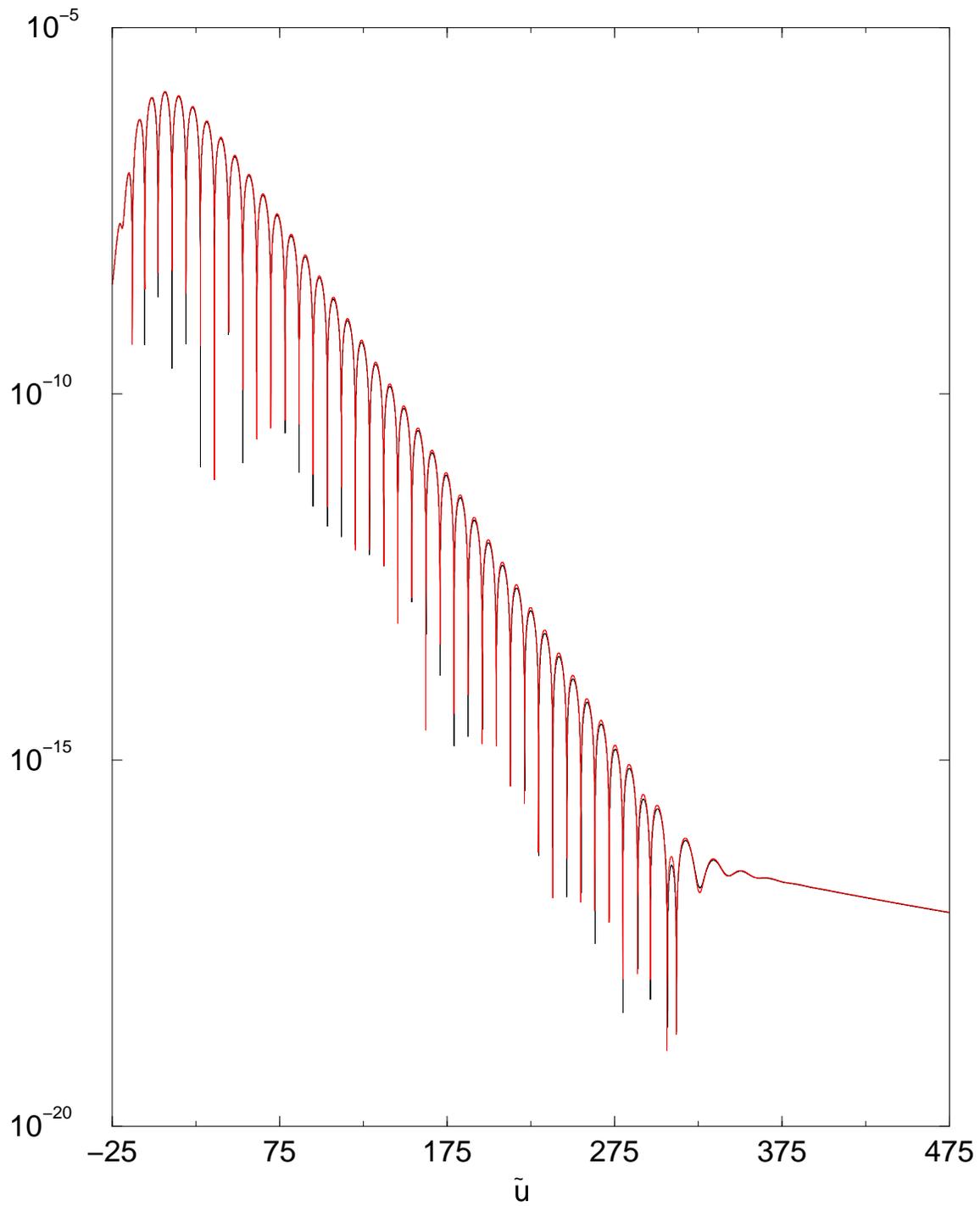}}
\caption{Convergence of the close approximation waveform at $\cal I^+$:
The overlaid plots of $4\delta y_1(\tilde u)$ and $\delta y_2(\tilde u)$ are
indistinguishable.}
\label{f_4_conformal_convergence_1}
\end{figure}

We test the convergence of the waveform at ${\cal I}^+$ by evolving this close
approximation data with increasingly larger grids containing 1001, 2001, and
4001 points. We define $\delta y_1$ to be the difference between the
waveforms obtained using 1001 and 2001 points, and $\delta y_2$ the difference
between using  2001 and 4001 points. (We consider only points common to all
three grids.) Second order convergence requires that $\delta y_1= 4\delta y_2$.
For these grid sizes, Figure \ref{f_4_conformal_convergence_1}  shows that
$\delta y_1$ and $4 \delta y_2$ overlap confirming that the code is second
order convergent throughout the quasinormal ringdown phase.

\begin{figure}
\centerline{\epsfxsize=6in\epsfbox{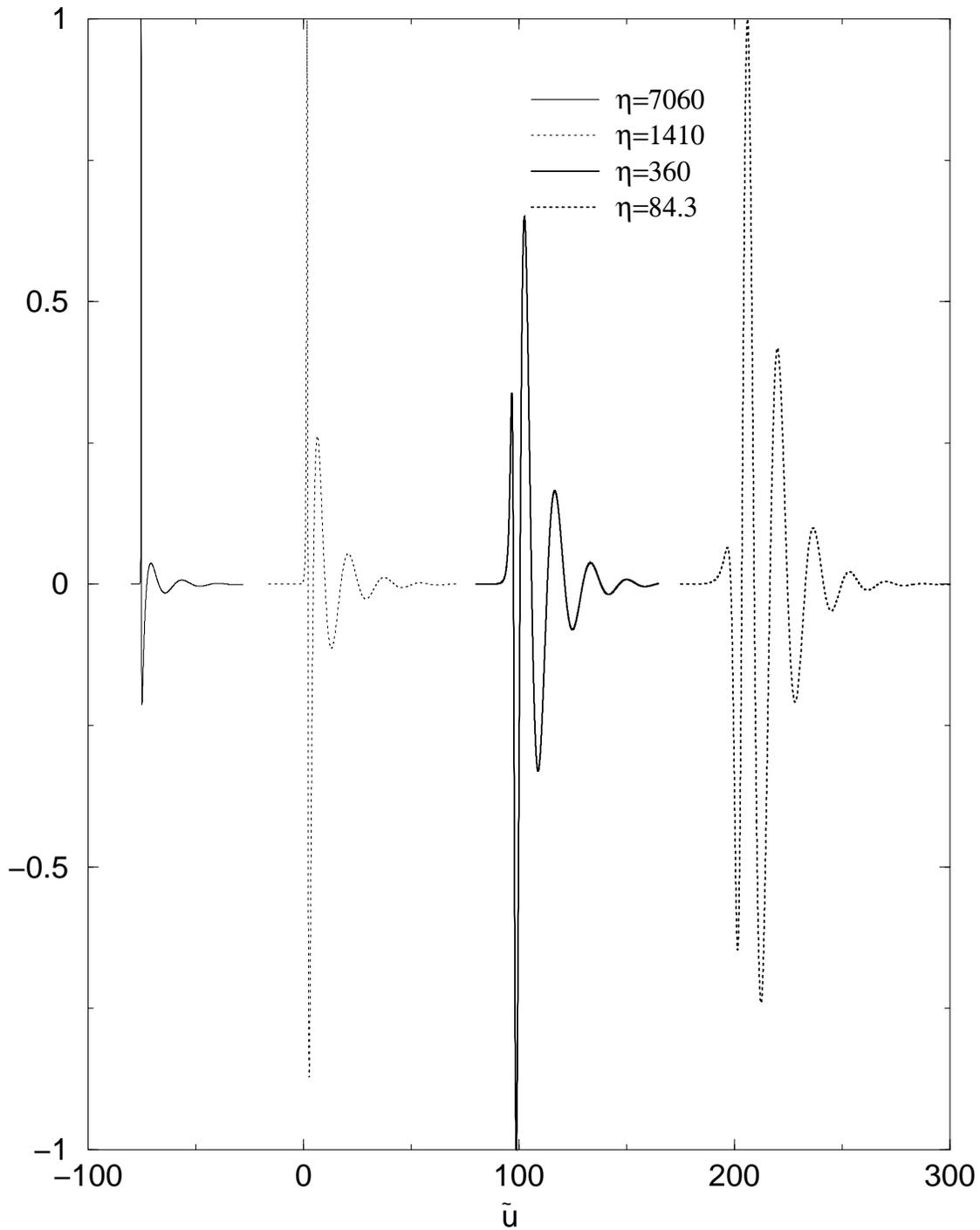}}
\caption{Close approximation waveforms ${\tilde F_4(\tilde u)}$ on $\cal I^+$
for $\eta = 7060$, $1400$, $368$ and $84.3$, with the amplitudes renormalized
by the relative factors of 1, 37.9, 544 and 11200, respectively.}
\label{F_4_conformal}
\end{figure}

Figure \ref{F_4_conformal} shows a series of waveforms produced on $\cal I^+$
obtained by evolving the close approximation data for  $\eta = 7060$, $1410$,
$364$ and $4.39$. The waveforms have been translated with respect to each
other and normalized to unit amplitude for purpose of comparison. The plots
show the waveforms from the initial time up to (roughly) the onset of
quasinormal decay.

\begin{figure}
\centerline{\epsfxsize=6in\epsfbox{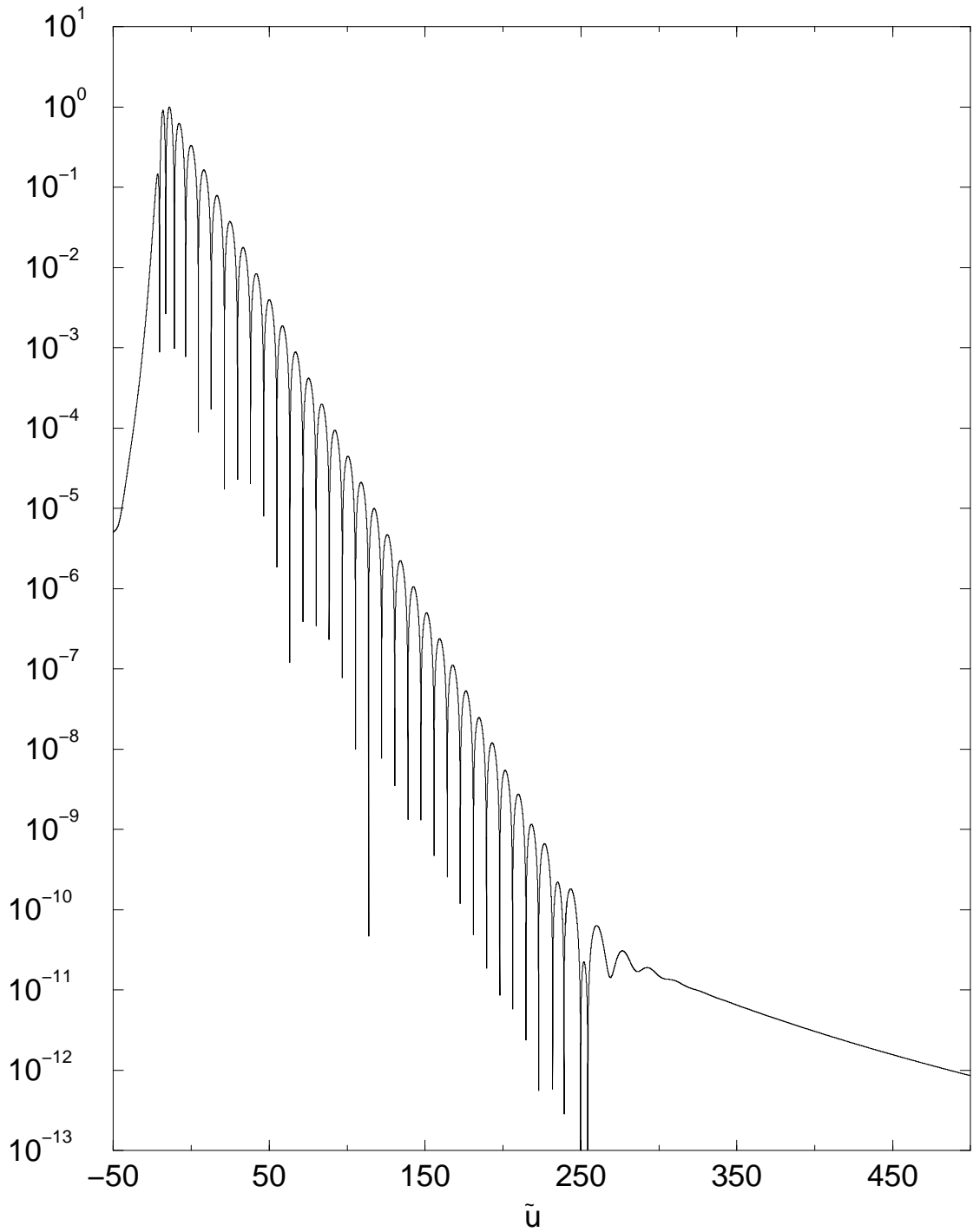}}
\caption{Close approximation waveform ${\tilde F_4(\tilde u)}$: Quasinormal
ringdown and tail for $\eta=158$. } \label{F_4_conformal_log} \end{figure}

Figure \ref{F_4_conformal_log} shows a log plot of the waveform
produced for $\eta=158$. The fit of the exponentially damped section  is
\begin{equation}
       \tilde F_4 \propto
           e^{-.0889575\tilde u }
	         \sin(.3736735\tilde u),
\end{equation}
which matches the the theoretical form for the lowest quasinormal mode
to five digits (in the frequency).

\begin{figure}
\centerline{\epsfxsize=6in\epsfbox{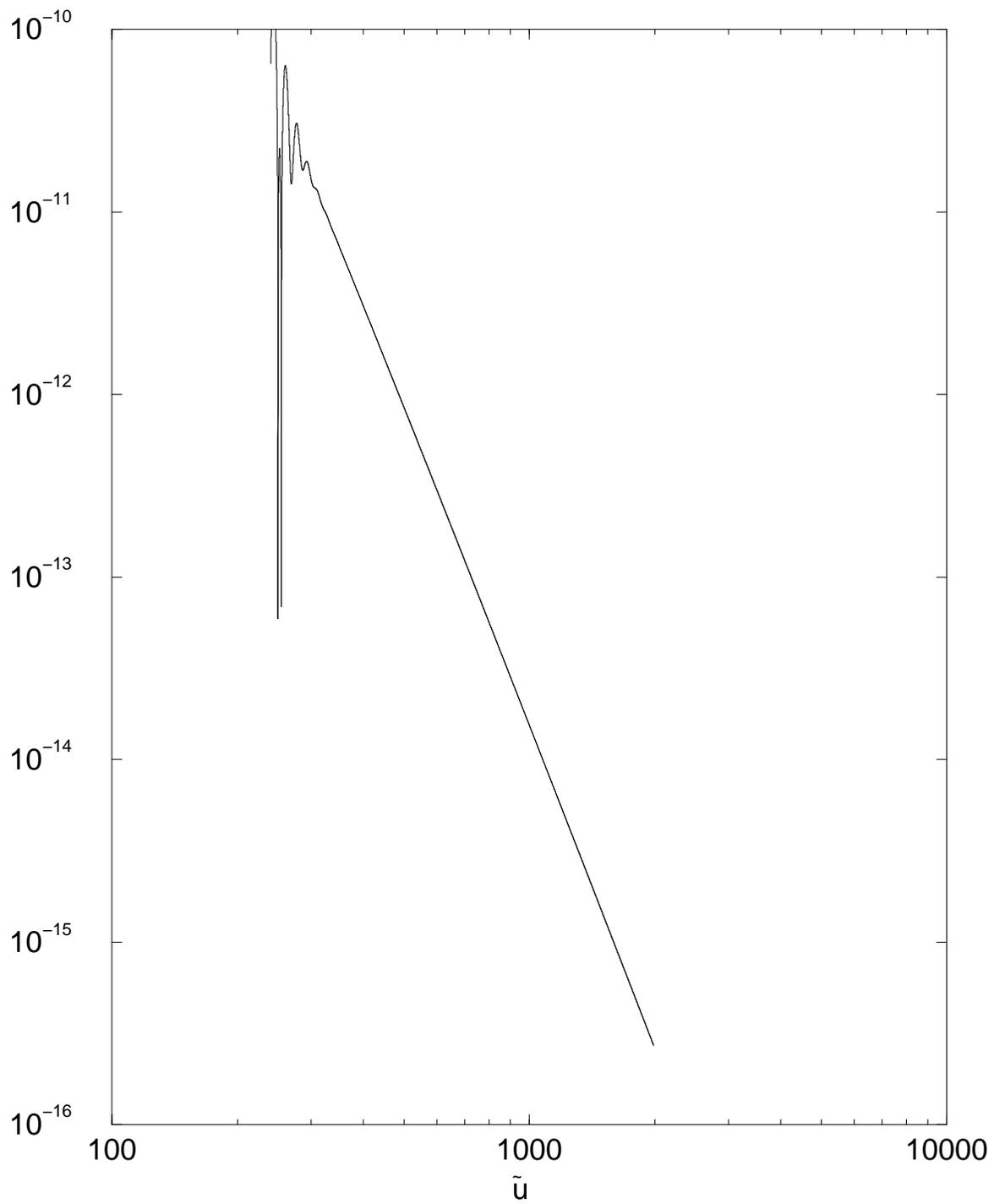}}
\caption{Close approximation waveform ${\tilde F_4(\tilde u)}$: Late time
power law tail for $\eta =158$.}
\label{F_4_conformal_tail}
\end{figure}

Figure \ref{F_4_conformal_tail} shows the late time tail of the waveform. The
measured slope of the  tail indicates a power law decay of the approximate form
$\tilde F_4 \propto \tilde u ^{-5.8}$ near the beginning of the tail to $\tilde
F_4 \propto \tilde u ^{-5.9}$ near the end, very similar to the behavior of the
tail for the compact pulse described in Sec. \ref{sec:compulse}. These results
suggest a final integer power law tail $\tilde F_4 \propto \tilde u ^{-6}$. For
an $\ell=2$ quadrupole wave, this is the same $\tilde u ^{-(2\ell +2)}$ integer
power law originally predicted by Price~\cite{price} for the decay of an
initially static multipole. A rigorous mathematical treatment of power law
tails has not yet been given~\cite{Kokkotas-Schmidt} and it would be
particularly interesting to reexamine the theory in the context of our boundary
conditions.

\section{Discussion}

Our results establish the capability of characteristic evolution of the
Teukolsky equation to determine an accurate advanced solution for the head-on
collision of black holes in the close approximation. In subsequent work, we
will extend these results to determine the physically more appropriate
retarded solution. In the fully nonlinear regime,  the conformal horizon model
for suppling binary black hole data, combined with an existing characteristic
evolution code, offers a new way to calculate the merger-ringdown waveform from
coalescing black holes. Because this is an unexplored area of binary black
hole physics, these perturbative studies of the head-on collision will provide
a preliminary physical check on extending the work to the nonlinear and
nonaxisymmetric case, where inspiraling black holes can be treated.

\acknowledgements

We thank Bernd Schmidt for numerous dicusssions. This work has been partially
supported by a Marie-Curie Fellowship (HPMF-CT-1999-00334) to M. C. and by NSF
grants PHY 9800731 and PHY 9988663 to the University of Pittsburgh. Computer
time for this project has been provided by the Pittsburgh Supercomputing Center
and by NPACI.


\end{document}